\documentclass[a4paper,11pt]{article}
\pdfoutput=1
\usepackage{jheppub}
\pdfoutput=1
\usepackage{epsf}
\usepackage{epsfig}
\usepackage{subfig}
\usepackage{mathtools}    
\usepackage{float}
\usepackage{multirow}
\usepackage{nicefrac}
\usepackage{epstopdf}
\usepackage{slashed}
\usepackage{xcolor}
\usepackage{url}
\usepackage{breqn}
\usepackage{amsmath}   
\usepackage{adjustbox} 
\usepackage{chemformula}
\usepackage{multicol}
\usepackage{placeins}
\usepackage{braket}
\usepackage{subdepth}
\usepackage{color}
\usepackage{booktabs}
\usepackage{siunitx}

\newcommand{\ifb}{\text{ fb}^{-1}}
\newcommand{\tev}{\text{TeV}}
\newcommand{\DSEWPO}{\text{OS}_{\text{EWPO}}}

\graphicspath{ {./Figures/} }

\title{{A global analysis of the SMEFT under the minimal MFV assumption}}

\author{Riccardo Bartocci, Anke Biekötter, Tobias Hurth} 

\affiliation{PRISMA+ Cluster of Excellence \& Institute of  Physics (THEP) \& Mainz Institute for Theoretical Physics, Johannes Gutenberg University, D-55099 Mainz, Germany}

\abstract{We present comprehensive global fits  of the SMEFT under the \textit{minimal} minimal flavour violation (MFV) hypothesis, i.e.~assuming that only the flavour-symmetric and CP-invariant operators are relevant at the high scale. 
The considered operator set is determined by theory rather than the used datasets. 
We establish global limits on these Wilson coefficients using leading order and next-to-leading order SMEFT predictions for electroweak precision observables, Higgs, top, flavour and dijet data as well as measurements from parity violation experiments and lepton scattering. 
Our investigations reveal an intriguing crosstalk among different observables, underscoring the importance of combining diverse observables from various energy scales in global SMEFT analyses.}

\preprint{
\begin{minipage}{3cm}
\small
\flushright
MITP/23-063
\end{minipage}}

\begin{document}

\maketitle

\section{Introduction}
\label{sec:introduction}

The Standard Model~(SM) of particle physics describes the interactions of fundamental particles with remarkable success. However, it fails to resolve some outstanding issues including the nature of dark matter, the origin of neutrino masses, and the electroweak hierarchy problem.
In the absence of a direct discovery of new physics~(NP), Standard Model Effective Field Theory (SMEFT)~\cite{Buchmuller:1985jz,Wilczek:1977pj,Grzadkowski:2010es,Brivio:2017vri} enables us to systematically explore the low-energy effects of NP at currently inaccessible energy scales.
SMEFT relies only on minimal assumptions and extends the dimension-four SM Lagrangian by higher-dimensional operators built from the SM fields and respecting the SM gauge symmetries.

The SMEFT framework has been extensively used to describe NP effects in experimental data and analyse the viable NP directions up to dimension six in global analyses. 
The selection of the Wilson-coefficient sets in these analyses has mostly been determined by the choice of the datasets and global analyses of low-energy~\cite{Falkowski:2017pss}, electroweak~\cite{Biekoetter:2018ypq,Kraml:2019sis,Dawson:2020oco,Almeida:2021asy,Anisha:2021hgc} and top~\cite{Buckley:2015lku,Aguilar-Saavedra:2018ksv,Brivio:2019ius,Bissmann:2019gfc,Durieux:2019rbz}  data as well as combinations thereof~\cite{Ellis:2020unq,Ethier:2021bye,Garosi:2023yxg} have been performed.
Nonetheless, global fits of all dimension-six SMEFT operators are (currently) intractable due to the enormous number of Wilson coefficients. 
The completely flavour-general SMEFT Lagrangian has 2499 free parameters, the so-called Wilson coefficients, at mass dimension six~\cite{Jenkins:2013zja,Jenkins:2013wua,Alonso:2013hga} (1350 of which are CP-even and 1149 are CP-odd). 
However, this number can be drastically reduced when considering specific flavour assumptions.

Since flavour observables push the appearance of flavour-violating operators far above the TeV scale~\cite{Calibbi:2017uvl,Silvestrini:2018dos}, it is natural to explore specific assumptions on flavour symmetries and their breaking pattern. These various symmetry assumptions lead to some model dependence and can be regarded as universality classes providing an organisation principle for the SMEFT, allowing to systematically analyse the pattern of NP~\cite{Faroughy:2020ina,Greljo:2022cah}.

Contributions to flavour-violating observables can be suppressed under the assumption of minimal flavour violation~(MFV), which allows the SM Yukawa couplings as the only sources of $U(3)^5$ breaking~\cite{Chivukula:1987py,Hall:1990ac,DAmbrosio:2002vsn}. 
Previous works within the standard MFV scenario restrict the number of operators to those relevant to the respective datasets considered in these analyses, see e.g.~\cite{Bruggisser:2021duo,Bruggisser:2022rhb,Grunwald:2023nli}. In the analyses of~\cite{Hurth:2019ula,Aoude:2020dwv} the so-called leading MFV hypothesis was used which includes the additional assumption that the NP at the high scale does not include any flavour changing neutral currents (FCNC)  at tree level. In all cases, the number of operators is finally determined by the considered datasets rather than by theory, as a global analysis of all NP directions in these scenarios is currently not possible\footnote{Setting only single-parameter limits on the Wilson coefficients in models under different flavour assumptions allows to set limits on the  general NP scale of such models, see e.g.\ recent analyses~\cite{Greljo:2023adz,Allwicher:2023shc}.}. 

In this work, we consider the \textit{minimal} MFV scenario and assume an exact $U(3)^5$ flavour symmetry of NP operators at the high scale within the MFV hypothesis. 
This assumption reduces the total number of Wilson coefficients to 47 at the high scale. Six of these coefficients correspond to CP-odd operators, which typically require dedicated observables to constrain them and therefore only have a minimal crosstalk with CP-even operators in global analyses~\cite{Ethier:2021ydt}, see e.g.~\cite{Ferreira:2016jea,Brehmer:2017lrt,Bernlochner:2018opw,Englert:2019xhk,Cirigliano:2019vfc,Biekotter:2020flu,Biekotter:2021int,Bakshi:2021ofj,Degrande:2021zpv,Bhardwaj:2021ujv,Hall:2022bme} for dedicated analyses of CP-odd operators.
We therefore set these contributions aside and consider a set of 41 (CP even) Wilson coefficients 
in total~\footnote{Starting from the 59 flavour-universal operators of the Warsaw basis~\cite{Grzadkowski:2010es} one finds that only 42 of them (36 CP even + 6 CP odd) can be made flavour symmetric without an additional Yukawa coupling. For 5 of the 36 CP-even operators, there are two independent flavour-symmetric ways to contract the flavour indices, which leads to 47 operators in total (with 41 CP even and 6 CP odd ones).}.
We denote this assumption the {\it minimal} MFV hypothesis because it corresponds to creating the minimal amount of flavour violation at the electroweak scale. 
In contrast to the MFV hypothesis, the exact $U(3)^5$ flavour symmetry assumed at the NP scale is not a renormalisation group invariant concept.

Our analysis confronts the Wilson coefficients of the minimal MFV SMEFT with data from electroweak precision observables (EWPO), low-energy parity violation experiments, as well as Higgs, top, flavour, Drell-Yan (DY) and dijet data. 
These datasets enable us to perform a complete global fit  within the minimal MFV hypothesis, testing a parameter set which is motivated by theory rather than data. 
We identify which datasets set the strongest bounds on each operator and analyse the crosstalk between different datasets in a global fit. 
For example, we will analyse the roles of Drell-Yan data as well as parity violation observables for constraining semileptonic operators and the interplay of top, flavour and dijet observables for constraining four-quark operators. 
Our main goal is to investigate the ingredients needed to perform a global fit of the SMEFT in which all the directions in parameter space are  reasonably constrained. We will show that in order to significantly constrain all 41 Wilson coefficients of the minimal MFV SMEFT, we have to rely on partial next-to-leading order (NLO) SMEFT predictions as some four-quark operators are essentially unconstrained at leading order (LO). 
The additional degeneracies induced in the SMEFT predictions at NLO only marginally impact the global analysis given the current level of constraints from other datasets.

Our paper is organised as follows: In Section~\ref{sec:symmetricSMEFT}, we define the minimal MFV and introduce the notation. In Section~\ref{sec:datasets}, we describe the considered datasets and list the Wilson coefficients constrained by each set. 
We present global analyses using LO predictions only or including partial NLO predictions in Sections~\ref{sec:LO_fit} and~\ref{sec:NLO_fit}, respectively. In Section~\ref{sec:interplay}, we analyse the interplay between different datasets in the global fit. 
We conclude in Section~\ref{sec:conclusions}.

\section{The $U(3)^5$ symmetric SMEFT }
\label{sec:symmetricSMEFT}

The SMEFT Lagrangian, truncated at dimension six, is given by 
\begin{equation}
\mathcal{L}_{\text{SMEFT}}=\mathcal{L}_{\text{SM}}+\sum_i \frac{C_i}{\Lambda^2} \, Q_i,
\label{eq:SMEFT}
\end{equation}
where $C_i$ are the Wilson coefficients of the operators $Q_i$ and $\Lambda$ denotes the NP scale, 
which we set to $\Lambda=4\,\tev$ throughout our work. 
We truncate all SMEFT predictions at linear order in the Wilson coefficients, neglecting quadratic contributions which are suppressed by $\Lambda^{-4}$ and therefore formally of the same order as dimension-eight contributions. We employ $\{\alpha,\,M_Z,\,G_F\}$ as our electroweak input parameters. 

Symmetries play a big role in the SM and beyond. 
The SM as well as the operators of SMEFT rely on Lorentz and gauge symmetries. 
A priori, there are no restrictions on the flavour structure in the SMEFT. However, the fact that explicit flavour violation is experimentally extremely constrained~\cite{Calibbi:2017uvl,Silvestrini:2018dos} makes it natural to assume symmetries for the flavour sector. 
We work in the up-basis, where the SM CKM matrix is located in the down component of the quark doublets, and assume no flavour violation at the NP scale in the SMEFT.  
Equivalently, we consider a $U(3)^5$ symmetry of the SMEFT, namely
\begin{equation}
U(3)^5=U(3)_\ell \times U(3)_q\times U(3)_e\times U(3)_u\times U(3)_d,
\end{equation}
where $\{\ell, q, e, u, d\}$ represent the SM fermions~\cite{Faroughy:2020ina}. 
The $U(3)^5$ symmetric SMEFT contains only the minimum and non-removable amount of flavour violation which results from the SM Yukawa couplings. 
Under this assumption, there are 47 independent SMEFT operators. 
Note that the contraction of the flavour indices of flavour-generic operators can result in two independent $U(3)^5$ singlets in some cases. This is the case for four-fermion operators with two fermion currents of the same chirality: $Q_{qq}^{(1)}$, $Q_{qq}^{(3)}$, $Q_{ll}$, $Q_{dd}$ and $Q_{uu}$.
For these operators, we hence introduce unprimed and primed Wilson coefficients which refer to the operators contracting the flavour indices within the fermion bilinears or between the two different fermion bilinears, respectively.  
As an example, the operator $Q_{ll} = (\bar l_i \gamma_\mu l_j)(\bar l_k \gamma^\mu l_l)$ has two flavour-symmetric contractions 
\begin{align}
C_{ll} \, \delta_{ij}\delta_{lk}  \qquad \text{and} \qquad  C_{ll}^{\prime} \, \delta_{ik}\delta_{jl} \, .
\end{align}
For the operator $Q_{ee}$, a Fierz identity implies a single independent coefficient $C_{ee}$. 
Limits on the primed and unprimed Wilson coefficients can generally be different. As an example, $C_{ll}^{\prime}$ enters the shift of the Higgs vacuum expectation value to leave the input parameter $G_F$ unchanged and is therefore tightly constrained through EWPO.  $C_{ll}$, on the other hand, is dominantly constrained through $e^-e^+\to l^-l^+$ observables and its limits are weaker than those on $C_{ll}^{\prime}$.

Note that even though all operators of the $U(3)^5$ symmetric SMEFT are flavour conserving, flavour-violating effects still occur due to flavour violation in the SM. The renormalisation group (RG) evolution from the NP scale $\Lambda$ down to the electroweak scale and below generates contributions to flavour-violating observables -- via the interplay between the flavour-violating SM interactions and the flavour-conserving NP vertices. Therefore, even the operators of the $U(3)^5$ symmetric SMEFT can be constrained by flavour observables~\cite{Hurth:2019ula}.

Six of the 47 $U(3)^5$ symmetric operators are (purely bosonic and) CP odd and are best constrained through dedicated CP-sensitive observables and have negligible crosstalk with CP even operators~\cite{Ferreira:2016jea}. Therefore, we set them aside in our analysis. 
We list the 41 CP even operators considered in our analysis in Table~\ref{tab:basis} in appendix~\ref{app:operators}.

\section{Datasets}
\label{sec:datasets}

\begin{figure}[thb]
    \centering
    \includegraphics[width=0.9\textwidth]{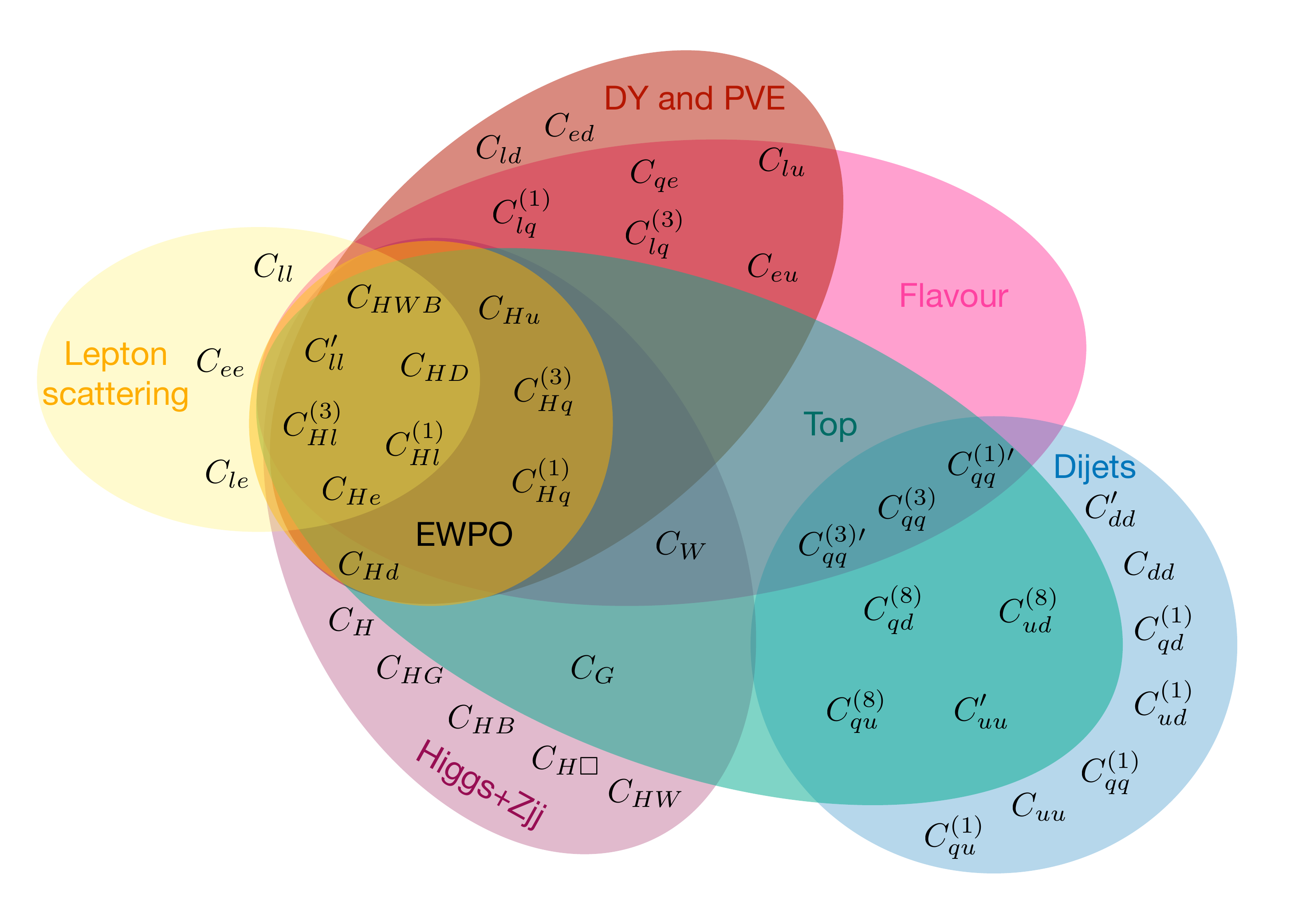}
    \caption{Venn diagram showing the operator sets contributing to the individual datasets at LO.}
    \label{fig:venn_diag}
\end{figure}

In our analysis, we include data from EWPO, Higgs, top, low-energy parity violation experiments (PVE), lepton scattering, flavour, Drell-Yan as well as dijet+photon production. 
We list the corresponding observables and operator sets constrained in each dataset in the following.
A graphical representation of operator sets contributing to the different observables is shown in Figure~\ref{fig:venn_diag}. 

\subsection{Electroweak precision observables}
We utilise a total of $13$ electroweak precision observables, including 
pseudo-measurements on the $Z$~resonance~\cite{ALEPH:2005ab}, a combination of the $W$~mass measurements at LEP~\cite{ALEPH:2013dgf}, Tevatron~\cite{Group:2012gb} and ATLAS~\cite{ATLAS:2017rzl} and the LEP and Tevatron combination of the decay width of the $W$~boson~\cite{ParticleDataGroup:2020ssz}\footnote{It was pointed out in~\cite{Berthier:2015gja,Berthier:2015oma,Bjorn:2016zlr,Berthier:2016tkq} that combinations of the measurements of $W$-boson mass and its decay width can be problematic as both results are partially based on the same datasets. We have explicitly checked that removing $\Gamma_W$ from our dataset does not influence our limits beyond the $5\%$ level. Moreover, it was suggested that LEP data should in principle not enter the combined measurement of the $W$ boson mass due to its extraction from $W^+W^-$ production, which makes it sensitive to modified triple-gauge couplings~\cite{Bjorn:2016zlr,Berthier:2016tkq}. However, given that the ATLAS and Tevatron measurements dominate the combination, we expect the impact on our fit to be small. }
\begin{equation}
\begin{split}
    \Gamma_{Z}, \;
    \sigma^{0}_{\text{had}}, \; 
    R^{0}_{l}, \; 
    A_{l}, \; 
   A^{l}_{FB}, \; 
    R^{0}_{c}, \;
    A_{c}, \; 
    A^{c}_{FB}, \;
    R^{0}_{b}, 
    A_{b}, \; 
    A^{0,b}_{FB}, \;
    m_{W}, \; 
    \Gamma_{W}  \, .
\end{split}
\end{equation}
These measurements constrain the following operator set at leading order
\begin{equation}
\DSEWPO = \{C_{HWB},	C_{HD},	C_{Hl}^{(1)},	C_{Hl}^{(3)},	C_{Hq}^{(1)},	C_{Hq}^{(3)},	C_{Hu},	C_{Hd},	C_{He},	C_{ll}^\prime\} \, .
\label{ops:EW}
\end{equation}

\subsection{Higgs and electroweak boson observables}
We include observables from the Higgs sector as included in~\cite{Anisha:2021hgc}. These include Higgs signal-strength measurements for various production and decay channels, simplified template cross section~(STXS) measurements and differential distributions. 
SMEFT predictions for the Higgs decays are based on~\cite{Brivio:2019myy}. 
For di-Higgs production, we include signal-strength measurements in the $4b$, $2b2\tau$ and $2b2\gamma$ final states. 
Moreover, we consider the angular distribution $\Delta \phi_{jj}$ in $Zjj$~production, which is known to dominantly constrain the Wilson coefficient $C_W$.
The full list of observables is given in Table~\ref{tab:obset}.
The set of operators contributing to Higgs observables and $Zjj$ production is given by 
\begin{equation}
\DSEWPO + \{C_W,  C_G, C_{HB}, C_{HW}, C_{HG}, C_{H}, C_{H\square}\} \, .
\label{ops:Higgs}
\end{equation}

\subsection{Top observables}
For the top sector, we reuse the datasets and corresponding predictions from fitmaker~\cite{Ellis:2020unq}. These include cross-section measurements, differential distributions and asymmetry observables for single-top, $t\Bar{t}$ and $ttV$ production.
The observables are listed in Tables~\ref{tab:obset_top}-\ref{tab:obset_top2_DY_dijet}. 
The set of operators contributing to top observables is given by 
\begin{equation}
\DSEWPO + \{C_W, C_G,  C_{qq}^{(1)\prime}, C_{qq}^{(3)}, C_{qq}^{(3)\prime},C_{uu}^{\prime},C_{ud}^{(8)},C_{qu}^{(8)},C_{qd}^{(8)}\} \, .
\label{ops:top}
\end{equation}

\subsection{Parity violation experiments and lepton scattering}
With the aim of constraining semileptonic operators, we include data from atomic parity violation (APV) and parity violating electron scattering (PVES) experiments, which we collectively refer to as parity violation experiments (PVE). 
Specifically, we include the weak charge $Q_W$ of \ch{^{133}Cs} \cite{ParticleDataGroup:2016lqr} and of the proton~\cite{Qweak:2013zxf}, deep inelastic scattering of polarised electrons as measured by the PVDIS experiment~\cite{PVDIS:2014cmd} and measurements of parity-violating scattering provided by SAMPLE~\cite{Beise:2004py}.

As a constraint on purely leptonic operators, we use muon neutrino-electron scattering, the weak mixing angle measured in parity violating electron scattering \cite{ParticleDataGroup:2016lqr}, $\tau$ polarisation measured in $e^+ e^-\to \tau^+\tau^-$~\cite{VENUS:1997cjg} as well as differential cross sections and asymmetries in $e^+ e^-\to l^+l^-$~\cite{VENUS:1997cjg,ALEPH:2013dgf,Electroweak:2003ram}.
The corresponding theory predictions are taken from~\cite{Falkowski:2017pss,Falkowski:2015krw}. 
The set of operators entering our low-energy observables is given by 
\begin{equation}
\DSEWPO + \{C_{lq}^{(1)}, C_{lq}^{(3)}, C_{ed}, C_{eu}, C_{ld}, C_{lu},  C_{qe}, C_{ll}, C_{ee}, C_{le}\} \, .
\label{ops:low_energy}
\end{equation}

\subsection{Drell-Yan observables}
Semileptonic operators can also be constrained in Drell-Yan lepton production. We use the \textit{HighPT}~\cite{Allwicher:2022mcg} tool to obtain the theory predictions for differential distributions of the $pp\to ee,\,\mu\mu,\tau\tau$ processes at leading order in the SMEFT. The corresponding experimental datasets are listed in Table~\ref{tab:obset_top2_DY_dijet}. 
To avoid conflicts with the EFT validity range, we only use invariant mass bins up to an energy of $3$~TeV. 
We describe the differential Drell-Yan cross sections by the following set of operators 
\begin{equation}
\DSEWPO (\text{without } C_{HWB}, C_{HD}, C_{ll}^\prime) + \{C_{lq}^{(1)}, C_{lq}^{(3)}, C_{ed}, C_{eu}, C_{ld}, C_{lu},   C_{qe} \}  \, .
\label{ops:DY}
\end{equation}
Note that currently shifts of the SM parameters as a result of the scheme choice are not taken into account in \textit{HighPT}. Consequently, the Wilson coefficients $C_{HWB}$, $C_{ll}^{\prime}$ and $C_{HD}$, which only contribute to the Drell-Yan dataset through these shifts, are not taken into account.
Given that these coefficients are tightly constrained by EWPO, we expect the impact of the missing contributions to be subdominant.

\subsection{Flavour observables}
We utilise the datasets and predictions as implemented in \textit{Flavio}~\cite{Straub:2018kue, Jenkins:2017jig,Jenkins:2017dyc,Dekens:2019ept} for flavour-sector observables.  
These include differential branching ratios of $B$~mesons and Kaons,  angular observables as well as the $R_K$ and $R_K^*$ ratios. A comprehensive list of the flavour observables included in our analysis is provided in Table~\ref{tab:obset_PVE_flavour}. 
As these observables are defined at low energies, we study them in the so-called Low Energy Effective Field Theory (LEFT). We present the effective Hamiltonians for the relevant flavour observables in Appendix~\ref{app:LEFT_Hamiltonians}. 
Starting from the Lagrangian definition in Equation~\eqref{eq:SMEFT}, we perform the running from the high scale ($\mu=4$~TeV) to the electroweak scale ($\mu=M_Z$), where we match the SMEFT onto the LEFT, using \textit{DsixTools}~\cite{Celis:2017hod,Fuentes-Martin:2020zaz,Jenkins:2017jig,Jenkins:2017dyc,Dekens:2019ept}. 
For the subsequent running from the electroweak~(EW) scale to the bottom-quark mass scale in the LEFT, we use the \textit{Wilson} package~\cite{Aebischer:2018bkb}. 
As a cross check, we have explicitly confirmed that we can reproduce the results presented in~\cite{Aoude:2020dwv}. 
The SMEFT Wilson coefficients appearing in the theory predictions of flavour violating processes are 
\begin{equation}
\DSEWPO (\text{without } C_{Hd}) + \{C_W, C_{lq}^{(1)},C_{lq}^{(3)},C_{lu},C_{eu},C_{qe}, C_{qq}^{(1)\prime}, C_{qq}^{(3)}, C_{qq}^{(3)\prime}\} \, . 
\label{ops:flavour}
\end{equation}

\subsection{Dijet+photon production}
LHC dijet production provides an interesting probe of the $4$-quark-operator parameter space.
However, the trigger threshold for jets at the LHC restricts the testable kinematic region to dijet invariant masses to the multi-TeV range. 
At high energies, the quadratic terms of the EFT expansion often dominate over the linear ones, potentially leading to conflicts with the EFT validity~\cite{Keilmann:2019cbp}.
Considering instead the production of two jets in association with a photon~\cite{ATLAS:2019itm} enables us to probe lower dijet invariant-mass ranges $m_{jj}< 1.2$~TeV and circumvent this issue. 

We have validated the SM dijet invariant-mass distribution of Figure~1 of~\cite{ATLAS:2019itm} using \textit{Madgraph}~\cite{MadGraph,Sjostrand:2014zea} Monte Carlo data and implemented the relevant cuts in a \textit{Rivet}~\cite{Buckley:2010ar} analysis for the event selection. We have excluded the region $m_{jj}< 500$~GeV, where we expect detector effects to be more relevant. 
Using a flat factor $\epsilon_{\text{det}}= 0.28$ for the experimental detector efficiency, we can reproduce the ATLAS SM distribution within~$10 \%$. 
Using SMEFTsim \cite{Brivio:2020onw}, we have determined the SMEFT prediction for each bin of the differential cross section at LO in SMEFT. The obtained predictions are available as an ancillary file with the arXiv submission. 

It is worth noting that some of the considered four-quark operators, $Q_{qd}^{(1)}$, $Q_{qu}^{(1)}$, $Q_{ud}^{(1)}$, do not interfere with the dominant SM diagram, the $t$-channel exchange of a gluon. 
As a result, we expect the quadratic SMEFT contributions for these operators, which we generally neglect in our study, to be dominant with respect to their linear counterparts.
To highlight this effect, we present the ratio for bounds from single-parameter fits based on linear SMEFT predictions only over bounds from linear+quadratic fits in Table~\ref{tab:dij-quad-single}. As anticipated, the limits on $C_{qd}^{(1)}$, $C_{qu}^{(1)}$, $C_{ud}^{(1)}$ are tightened by more than a factor $2.5$ in a quadratic fit. Since we will only underestimate the limits on these Wilson coefficients in our linear fit, we will keep their linear contributions. The remaining operators are much more mildly influenced, typically by $10-20\%$, in a quadratic fit. 

{
\setlength{\tabcolsep}{5pt}
\begin{table}[t]
\centering
\begin{tabular}{ccccccccccccccc}
\hline
$C_{qq}^{(1)}$ &  $C_{qq}^{(1)\prime}$ &$C_{qq}^{(3)}$ & $C_{qq}^{(3)\prime}$ &$C_{qd}^{(8)}$ & $C_{qu}^{(8)}$ &$C_{ud}^{(8)}$ &$C_{dd}$ &$C_{dd}^{\prime}$ & $C_{uu}$ & $C_{uu}^{\prime}$ & $C_{qd}^{(1)}$ & $C_{qu}^{(1)}$ &  $C_{ud}^{(1)}$ &  $C_{G}$\\
\hline
0.9 & 0.9 & 1.0 & 1.0 & 1.7 & 1.4 & 1.6 & 2.5 & 2.2 & 0.9 & 0.9 & 5.9 & 3.9 & 2.6 & 1.4 \\  
\hline
\end{tabular}
\caption{Ratio of the $68\%$ CL limits on the operators contributing to dijet+photon production at linear and quadratic order in SMEFT. }
    \label{tab:dij-quad-single}
\end{table}
}

We neglect the clearly subdominant electroweak contributions to dijet production and describe the differential distributions by 
\begin{equation}
\{ C_G , C_{qd}^{(1)}  ,   C_{qu}^{(1)}, C_{ud}^{(1)},  
C_{qq}^{(1)}, C_{qq}^{(1)\prime}, C_{qq}^{(3)},C_{qq}^{(3)\prime}, 
C_{uu},  C_{uu}^{\prime}, C_{dd},  C_{dd}^{\prime}, 
C_{ud}^{(8)}, C_{qu}^{(8)}, C_{qd}^{(8)} \}  \, .
\label{ops:dijet}
\end{equation}

\section{Global fit using leading order predictions}
\label{sec:LO_fit}
%
We perform a $\chi^2$ analysis of the data $\Vec{d}$ with linear SMEFT predictions $\Vec{p}(C_i)$ taking into account correlations in the covariance matrix $V$, where known. The $\chi^2$ function is given by
\begin{align}
    \chi^2(C_i) = \left( \Vec{d} - \Vec{p} (C_i) \right)^T V^{-1} \left( \Vec{d} - \Vec{p}(C_i) \right) \, .
\end{align}
For our global analyses, we obtain the limits on a Wilson coefficient while profiling over the remaining parameters using the toy Monte Carlo method. 

\begin{figure}[thb]
    \centering
    \includegraphics[width=\textwidth]{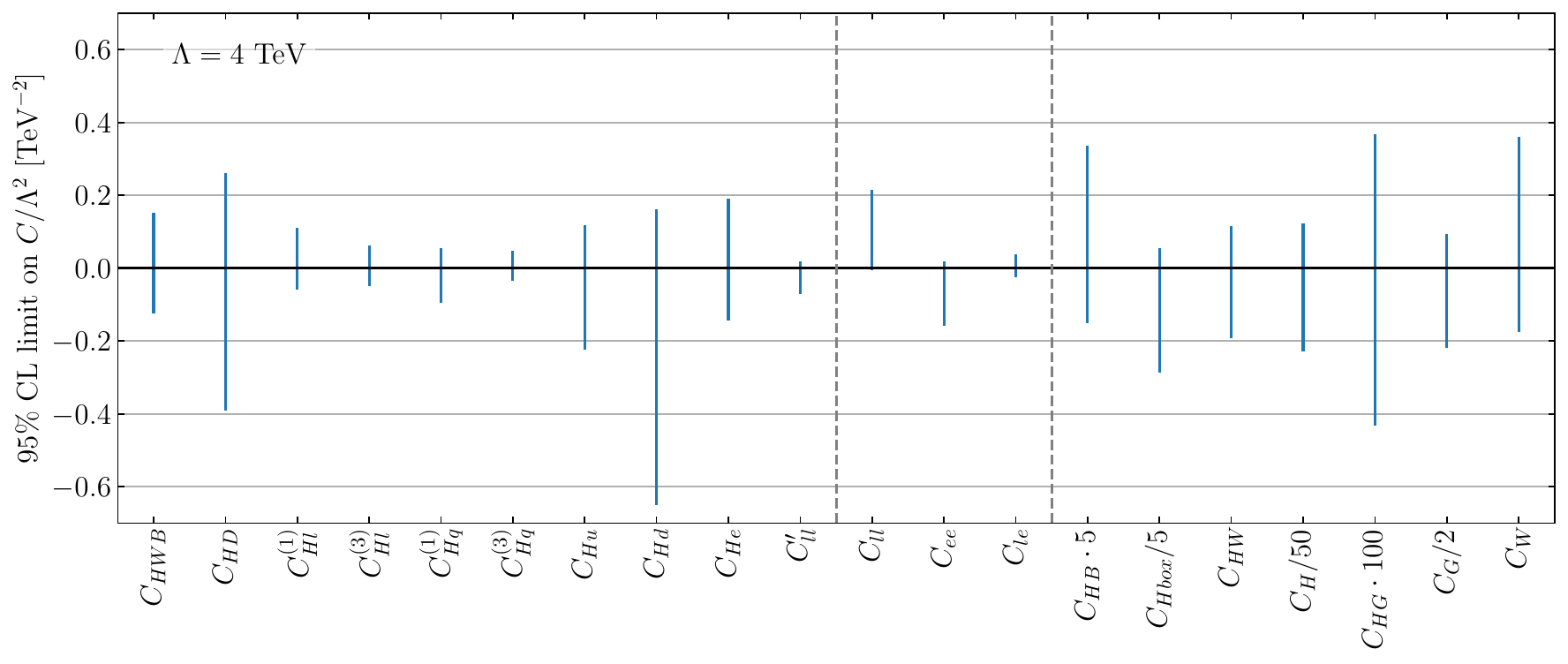}
    \includegraphics[width=\textwidth]{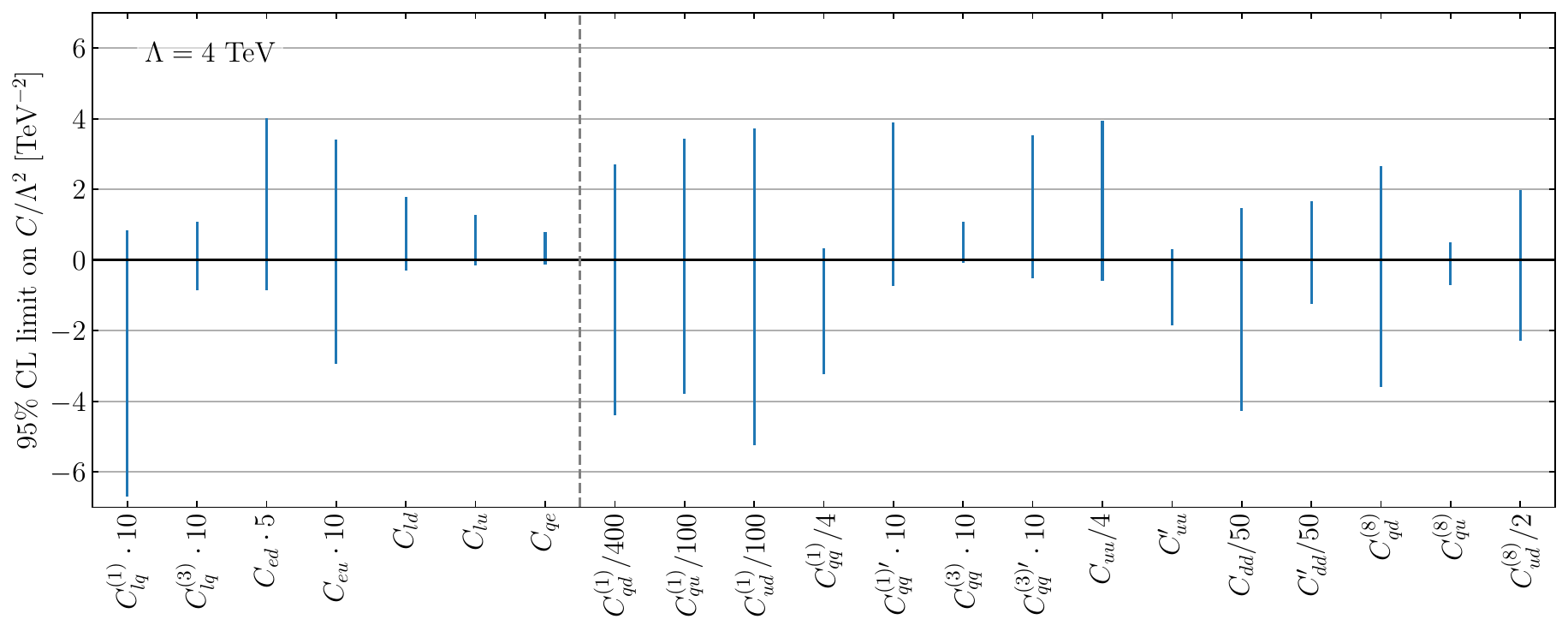}
    \caption{Limits on the Wilson coefficients in a global analysis using LO SMEFT predictions.}
    \label{fig:global_LO} 
\end{figure}

We present our global analysis  with SMEFT predictions at LO in Figure~\ref{fig:global_LO}. Operators from the Higgs-EW sector as well as four-lepton operators, shown in the upper panel of the plot, are well constrained. All corresponding Wilson coefficients, except $C_H$, are constrained to a region within $|C|/\Lambda^2 < 1/\text{TeV}^2$ at $95\%$ CL. 
The four-fermion operators involving quark fields, shown in the lower panel of the plot, are generally more weakly constrained. The limits of two of the seven semileptonic operators as well as nine of the 14 four-quark operators exceed $|C|/\Lambda^2 = 1/\text{TeV}^2$ (on at least one side). 
In particular, $C_{qd}^{(1)}$, $C_{ud}^{(1)}$, $C_{qu}^{(1)}$, $C_{dd}$ and $C_{dd}^{\prime}$ are essentially unconstrained, with limits on $|C|/\Lambda^2$ greater than $50/\text{TeV}^2$. Nevertheless, it is interesting to note that the inclusion of these very weakly constrained operators does not invalidate the limits on the remaining Wilson coefficients.

All 41 Wilson coefficients included in our global fit are consistent with the SM within~$2\sigma$. However, six Wilson coefficients exhibit deviations exceeding $1.5\sigma$
\begin{align}
\label{eq:dev_1p5sigma_LO}
\{ C_{ll}, \,  C_{ee}, \, C_{lu}, \, C_{lq}^{(1)}, \, C_{qq}^{(1)}, \, C_{qq}^{(3)}\} \, .
\end{align}
The shift in $C_{ll}$ can be attributed mainly to a small deviation in the measurement of the differential cross-section of $e^+e^-\to e^+e^-$, while $C_{ee}$ is affected by the measurement of the weak mixing angle in parity-violating electron scattering $\left(g_{AV}^{ee}\right)$.
Several semileptonic operators exhibit deviations from the SM, with the largest shifts arising for $C_{lu}$ and $C_{lq}^{(1)}$. These deviations are primarily induced by the Drell-Yan data, in particular with muons in the final state. 
In the four-quark sector, $C_{qq}^{(3)}$ is dominantly shifted by single-top production data and the shift of $C_{qq}^{(1)}$ is dominated by dijet data, which is the only observable constraining this operator at LO.

\section{Global fit including next-to-leading order predictions}
\label{sec:NLO_fit}

As shown in the previous section, some four-quark operators within the minimal MFV set remain weakly constrained in a leading-order fit to our datasets. These operators contribute to a number of precisely measured observables at NLO. 
In this section, we perform a global analysis based on NLO predictions, where present. This includes NLO predictions for EWPO~\cite{Dawson:2019clf,Dawson:2022bxd} as well as partial NLO predictions for $t\Bar{t}$ production~\cite{Hartland:2019bjb,Kassabov:2023hbm} and Higgs decays to bottom quark pairs~\cite{Alasfar:2022zyr}. 
Note that not all observables are considered at NLO precision in our partial NLO global analysis. As a result, our analysis does not consider all degeneracies which might be present in the SMEFT predictions at NLO\footnote{See e.g.~\cite{Alasfar:2022zyr} for an example of how the consideration of further NLO effects can spoil the loop sensitivity to the Higgs self-coupling.}. 
Nevertheless, the inclusion of EWPO at NLO precision serves as a good testcase to study the impact of the additional freedom in the Wilson coefficient space at NLO on the operators present already at LO.

\subsection{SMEFT predictions at NLO}

\paragraph{EWPO}
We include NLO predictions for EWPO from~\cite{Dawson:2019clf,Dawson:2022bxd}, see also~\cite{Biekotter:2023xle} for partial results in other EW input schemes. 
At NLO precision, EWPO are sensitive to 25 additional Wilson coefficients. Explicitly, the considered NLO EWPO predictions are sensitive to 
\begin{align}
    \{ & 
    C_{ll} , \, C_{ee} , \, C_{le} , \, 
    C_{HB} , \, C_{H \square} , \, C_{HW}, \, C_W , \, 
    C_{lq}^{(1)}, \, C_{lq}^{(3)}, \, C_{ed} , \, C_{eu} , \, C_{ld}, \, C_{lu}, \, C_{qe},  
    \nonumber \\ & 
    C_{qd}^{(1)}, \, C_{qu}^{(1)} , \, C_{ud}^{(1)} , \,
    C_{qq}^{(1)} , \, C_{qq}^{(1)\prime} , \, C_{qq}^{(3)}, \, C_{qq}^{(3)\prime} , \, 
    C_{uu} , \, C_{uu}^{\prime} , \, C_{dd} , \, C_{dd}^{\prime} 
    \} \, ,
\end{align}
in addition to those listed in Equation~\eqref{ops:EW}.

\paragraph{Top} 

SMEFT predictions for top observables at NLO are provided by the SMEFit collaboration~\cite{Hartland:2019bjb,Kassabov:2023hbm}, which employs the $\{G_F, \, M_W, \, M_Z\}$ input scheme. While this input scheme is different from the one utilised in our analysis, the impact of the electroweak input-scheme choice on top quark physics is expected to be small. We have explicitly checked that the LO predictions for $m_{tt}$ differential distributions are very similar in the two input schemes. 
To improve the bounds on some four-quark operators, we update the predictions for the charge asymmetry as well as $m_{tt}$ differential distributions to NLO precision as a proof-of-principle. These observables have been shown to have the largest constraining power for four-quark operators~\cite{Kassabov:2023hbm}. 
NLO predictions of the considered top quark production processes add sensitivity to the operators
\begin{equation}
    \{ C_{qd}^{(1)} , \, C_{qu}^{(1)} , \, C_{ud}^{(1)}, \, C_{qq}^{(1)} , \, C_{uu} \} \, .
\end{equation}

\paragraph{Higgs} 

The dominant contributions of third generation four-quark operators to single-Higgs production and decay are known~\cite{Alasfar:2022zyr}. We include the NLO predictions for gluon fusion Higgs production, $t\bar{t}h$ production and the loop-induced decays $h \to g g , \, \gamma \gamma$, which receive contributions from the operators 
\begin{equation}
    \{ C_{qu}^{(1)}, \, C_{qq}^{(1)} , \, C_{qq}^{(3)} , \, C_{uu} , \,  C_{qu}^{(8)} \} \, .
\end{equation}
Specifically, $C_{qu}^{(1)}, \,C_{uu} , \,  C_{qu}^{(8)}$ contribute to $t\bar{t}h$ production only, while the remaining operators affect all considered processes. 
While the constraining power of Higgs data for these operators, which appear at NLO only, is generally small, we find that in particular the bounds on $C_{qu}^{(1)}$ tighten with the inclusion of these NLO effects. 

Moreover, we include NLO contributions from $C_H$ to total Higgs production and decay channels~\cite{Degrassi:2016wml,Degrassi:2019yix}. However, this only marginally affects the results as the constraints on $C_H$ are dominated by di-Higgs production.

\subsection{Global fit}

\begin{figure}[thb]
    \centering
    \includegraphics[width=\textwidth]{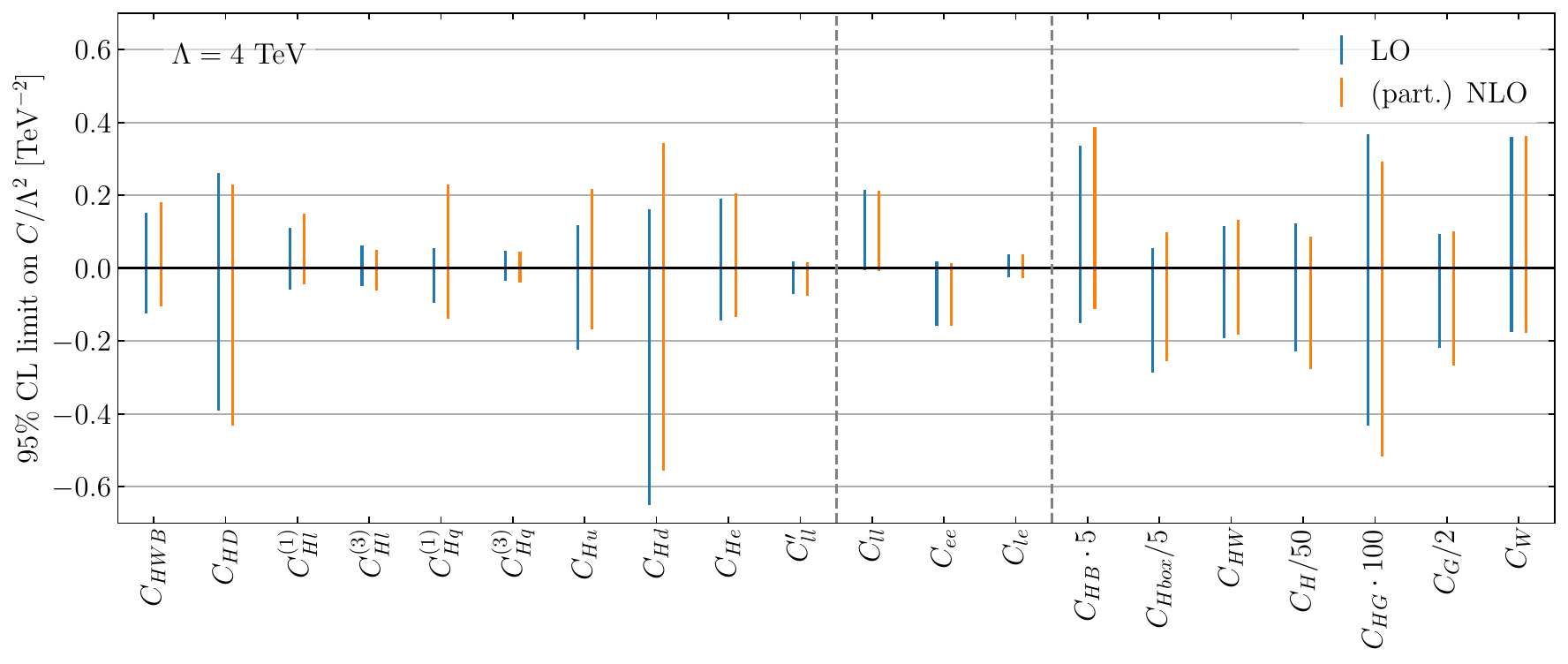}
    \includegraphics[width=\textwidth]{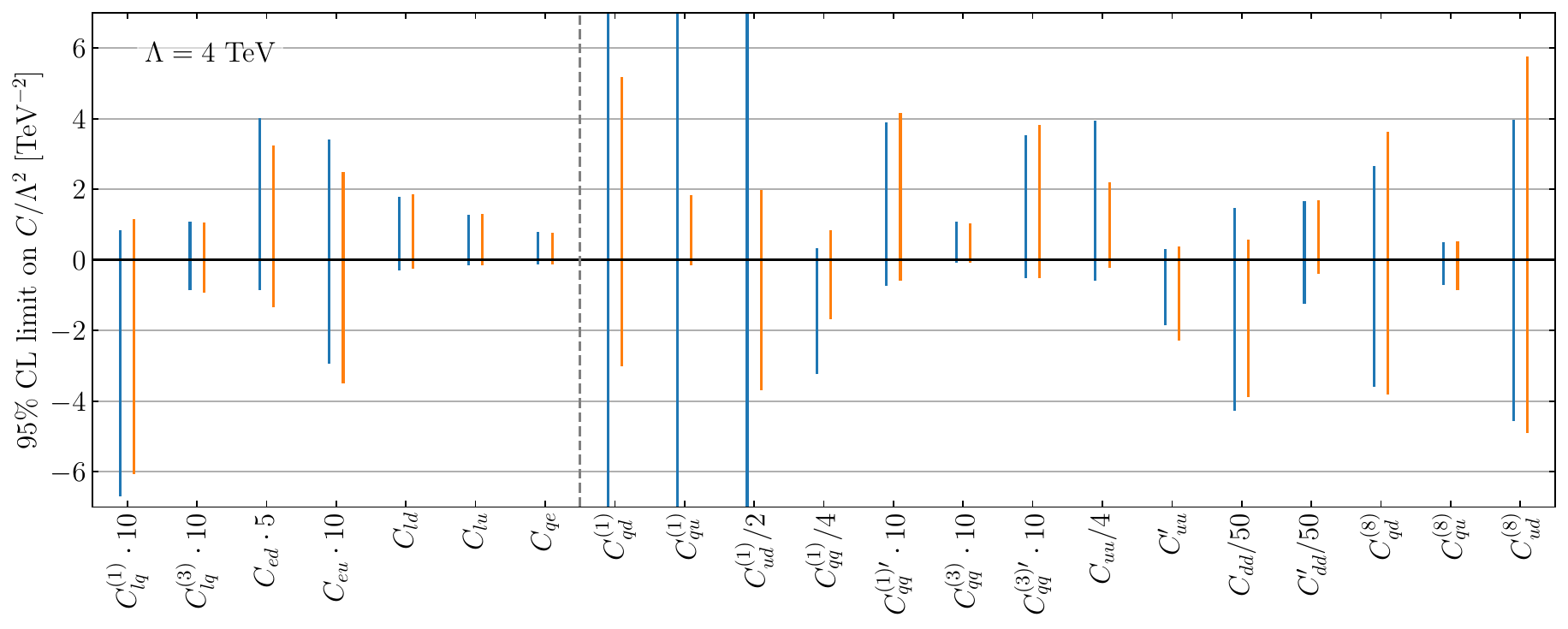}
    \caption{Comparison of the global analysis at LO with the one including partial NLO predictions.}
    \label{fig:LO_vs_NLO_fit}
\end{figure}

\begin{figure}
    \centering
     \includegraphics[width=0.45\textwidth]{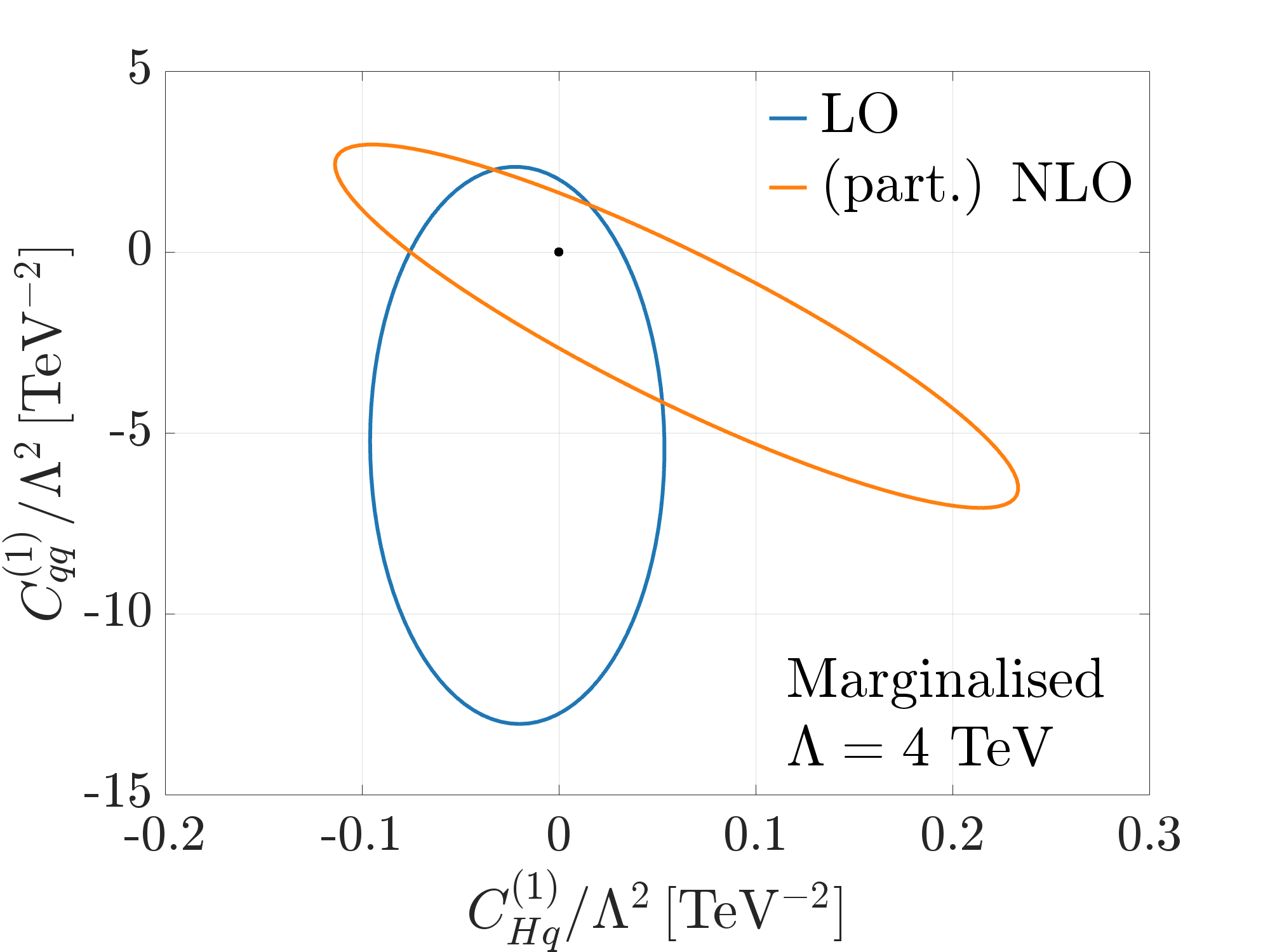}
    \includegraphics[width=0.45\textwidth]{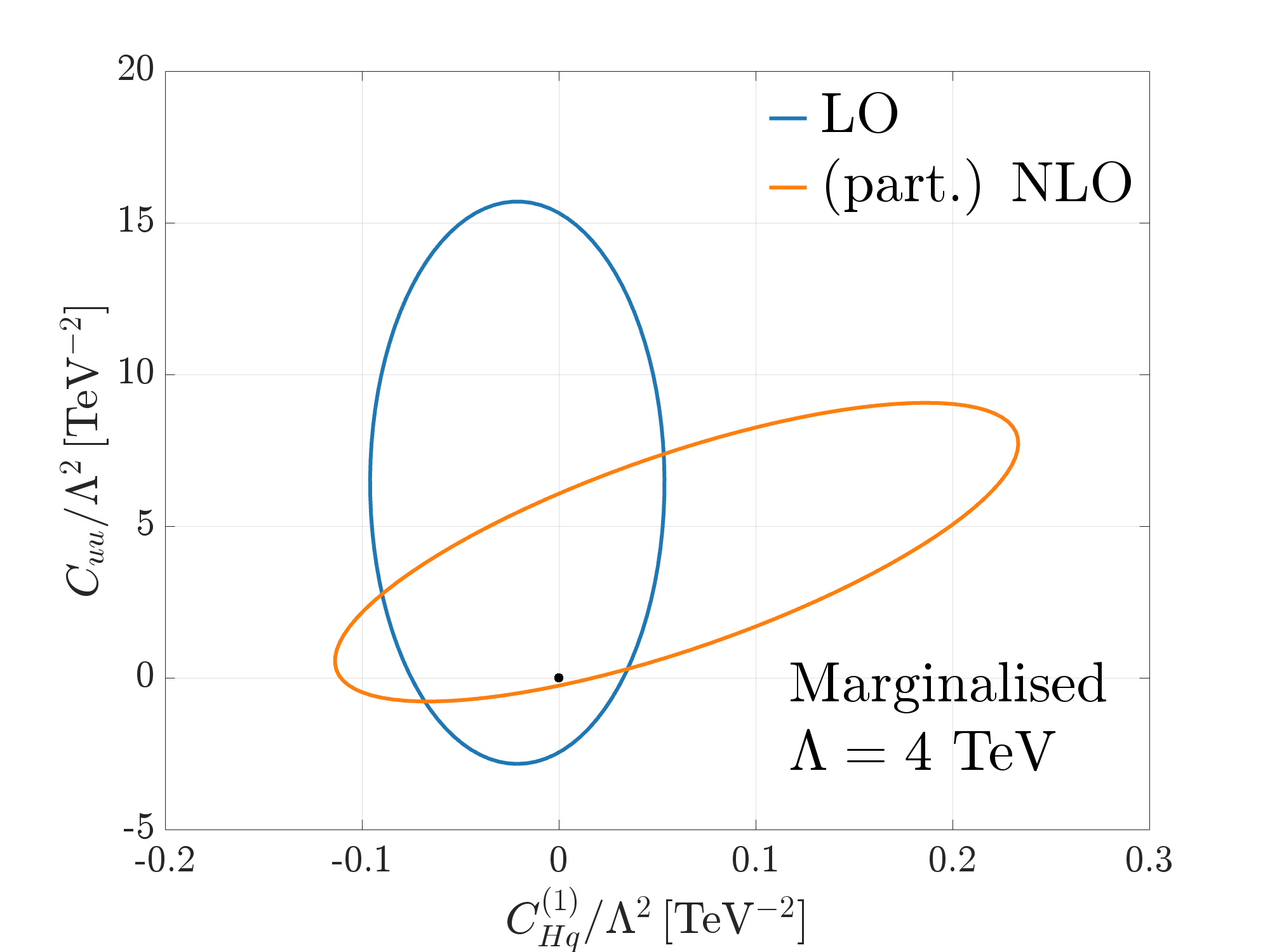}
    \caption{95\% CI limits in the LO and partial NLO analyses showing the correlations of $C_{Hq}^{(1)}$ with $C_{qq}^{(1)}$ (left) and $C_{uu}$ (right).}
    \label{fig:NLO_CHq1}
\end{figure}

In Figure~\ref{fig:LO_vs_NLO_fit}, we present a comparison of the global fit at LO with the one including partial NLO SMEFT predictions. The corresponding numerical fit results can be found in Table~\ref{tab:fits_numerical} in appendix~\ref{app:num_res}.  The bounds on most operators are only mildly influenced by the inclusion of NLO SMEFT predictions. 
However, in the four-quark sector the limits on the Wilson coefficients $C_{qd}^{(1)}$, $C_{qu}^{(1)}$, $C_{ud}^{(1)}$ and, to a smaller extent, $C_{uu}$ significantly tighten when improving the SMEFT predictions to NLO precision and are now below $|C|/\Lambda^2 < 10/\text{TeV}^2$. The Wilson coefficients $C_{dd}$ and $C_{dd}^\prime$ remain the only ones exceeding this limit. 
On the other hand, the bounds on $C_{Hq}^{(1)}$ are weakened by a factor 2.5 through its correlations with four-quark operators in EWPO, in particular with $C_{qq}^{(1)}$ and $C_{uu}$. We show these correlations arising at NLO in Figure~\ref{fig:NLO_CHq1}. 
At LO, there are no visible correlations between $C_{Hq}^{(1)}$ and the four-quark operators as indicated by the blue contours. At NLO, shown in orange, a strong correlation with $C_{qq}^{(1)}$ and an anti-correlation with $C_{uu}$ are induced through the EWPO SMEFT predictions, weakening the limits on $C_{Hq}^{(1)}$.

All 41 Wilson coefficients remain consistent with the SM within~$2\sigma$ in the NLO fit. However, the number of Wilson coefficients exhibiting deviations exceeding $1.5\sigma$ grows from six to nine
\begin{align}
\{ {\color{gray}C_{ll}, \, C_{ee}, \, C_{lu}, \, C_{qq}^{(3)} } , \, C_{ld}, \, C_{qu}^{(1)}, \, C_{qq}^{(1\prime)}, \, C_{qq}^{(3\prime)} , C_{uu} \} \, ,
\end{align}
where we have greyed out the Wilson coefficients which already deviated from zero in the LO fit, see Equation~\eqref{eq:dev_1p5sigma_LO}.
Two coefficients, which exhibit deviations $\geq 1.5\sigma$ in the LO fit are shifted towards more SM-like values at NLO, $C_{qq}^{(1)}$ and $C_{lq}^{(1)}$. For $C_{qq}^{(1)}$, this is due to its additional NLO contributions to EWPO, Higgs and top data. The shift of $C_{lq}^{(1)}$ towards more SM-like values in the NLO fit is the result of EWPO data.
On the other hand, $C_{qu}^{(1)}$, which at NLO is strongly constrained by Higgs data, deviates due to experimental deviations in $ttH$ production. 
$C_{uu}$ experiences a shift away from zero in both the LO and NLO analyses, which is caused by the dijet dataset. However, the shift becomes more apparent at NLO, where the overall bounds on this parameter shrink as a result of stronger limits on other correlated four-quark operators. 
For $C_{qq}^{(1)\prime}$, which is dominantly affected by $t\bar{t}V$, and $C_{qq}^{(3)\prime}$, which is mainly shifted by both $t\bar{t}V$ and dijet data, the change of the limits between the LO and NLO analysis is only marginal, the central values are shifted by $13 \%$ and $9\%$, respectively.
The same is true for the semileptonic operator $C_{ld}$, for which the central value only changes by $8\%$.

We have again checked explicitly that when conducting a fit without the five most poorly constrained operators, $C_{qd}^{(1)}$, $C_{ud}^{(1)}$, $C_{qu}^{(1)}$, $C_{dd}$ and $C_{dd}^{\prime}$, limits on the remaining operators exhibit no significant changes. 
The only noticeable effects are on four-quark operators for which the corresponding limits strengthen by up to $33 \%$ when removing the five least constrained operators from the fit. Moreover, the limit on $C_G$ changes by $17\%$ and the limit on $C_{Hq}^{(1)}$, which suffers from large correlations in the global analysis, receives a $35\%$ correction. 
This highlights the robustness of the results even though some operators remain weakly limited given the current dataset.

\begin{figure}[thb]
    \centering
    \includegraphics[width=\textwidth]{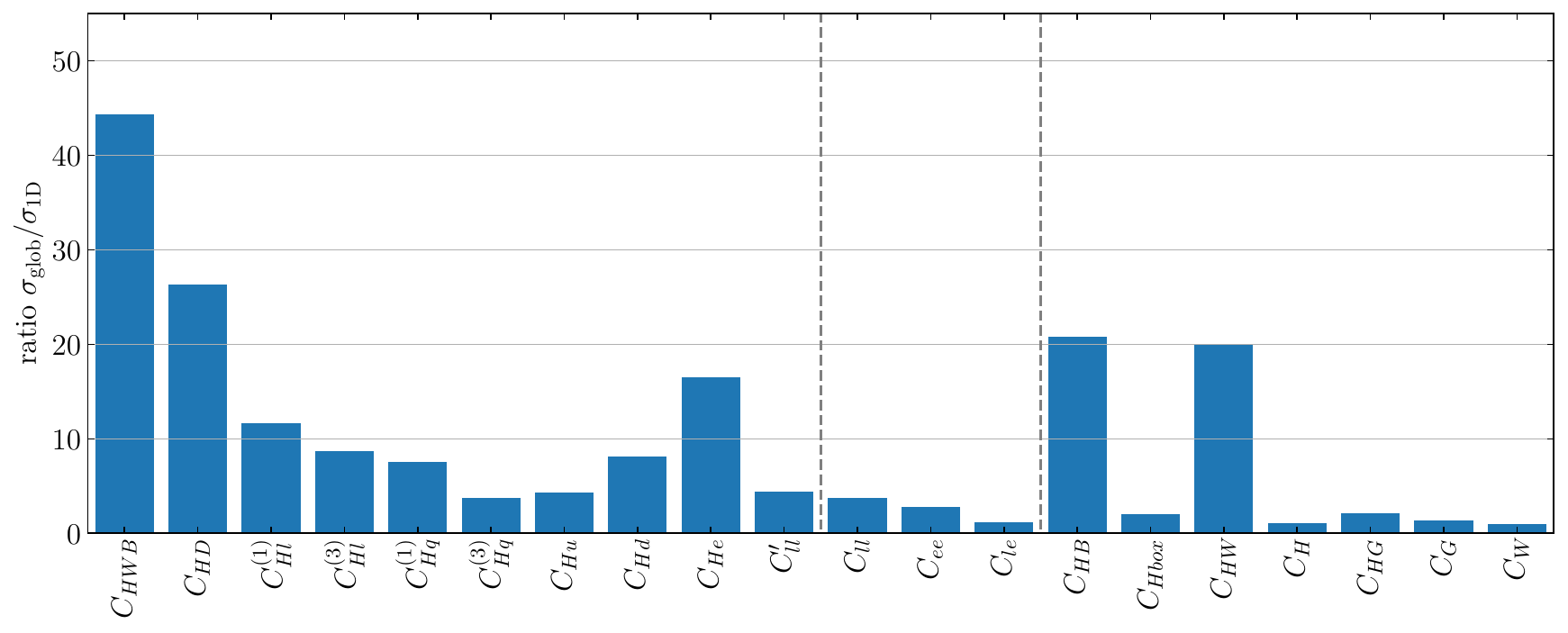}
    \includegraphics[width=\textwidth]{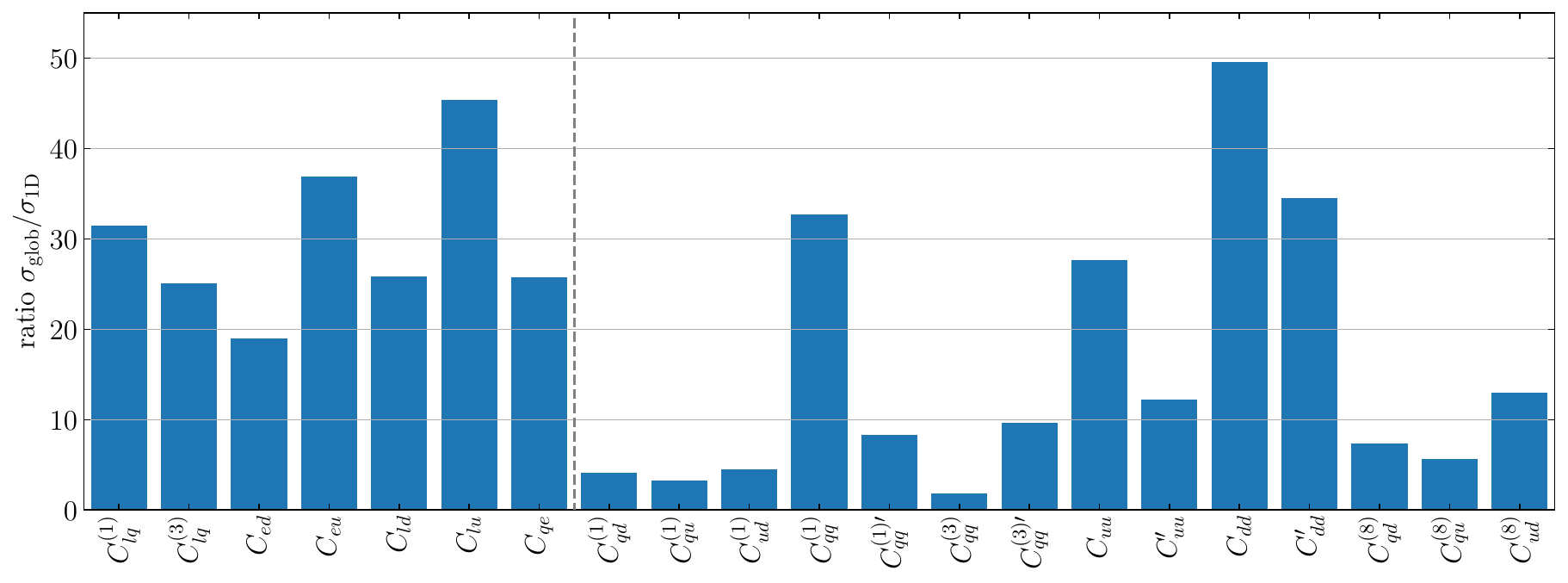}
    \caption{Ratio of the uncertainties on the Wilson coefficients in a global 41-parameter analysis over the uncertainties in a single-parameter analysis (both including partial NLO predictions). This can be interpreted as a measure of the relevance of correlations with the remaining Wilson coefficients in the analysis.}
    \label{fig:ratio_global_1D}
\end{figure}

In Figure~\ref{fig:ratio_global_1D} we present the ratio of the uncertainties on the Wilson coefficients in a global fit over the uncertainties in a one-parameter fit. 
This can be seen as a measure of the sensitivity of each operator on correlations with the remaining ones. 
As expected, we find that the bounds on many operators are significantly weakened in the global analysis. Limits on semileptonic operators and four-quark operators involving right-handed quark fields only, typically increase by more than a factor 20 in a global fit. 
For semileptonic operators, this is due to the Drell-Yan dataset imposing tight constraints at the one-parameter level, while only poorly disentangling the effects of different operators. We will further discuss this in Section~\ref{sec:DY_PVE}. 
Of those operators affecting EWPO at LO, $C_{HWB}$ and $C_{HD}$ have the largest degeneracies and the corresponding limits increase by factors 44 and 26, respectively, in a global fit. 
In the Higgs-gauge sector, only $C_{HW}$ and $C_{HB}$, which have  large correlations with $C_{HWB}$ and $C_{HD}$, see also the correlation matrix in Figure~\ref{fig:corr_mat}, are significantly weakened in the global analysis by factors of 20 and 21, respectively. 
The operators whose limits are most stable under a global analysis are $C_{le}$, $C_H$, $C_G$ and $C_W$. The corresponding limits change by less than 50~\% in a global fit.

\section{Interplay between the datasets}
\label{sec:interplay}
\begin{figure}
    \includegraphics[height=6.0cm]{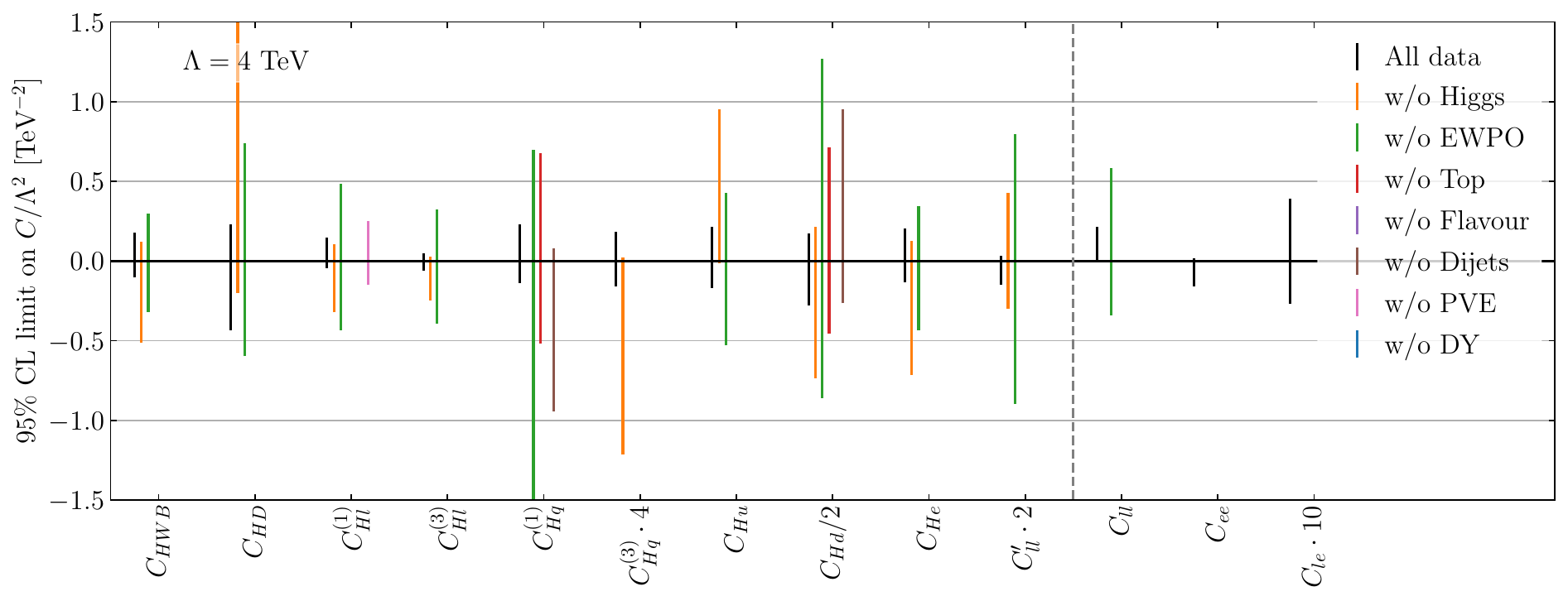}
    \includegraphics[height=6.0cm]{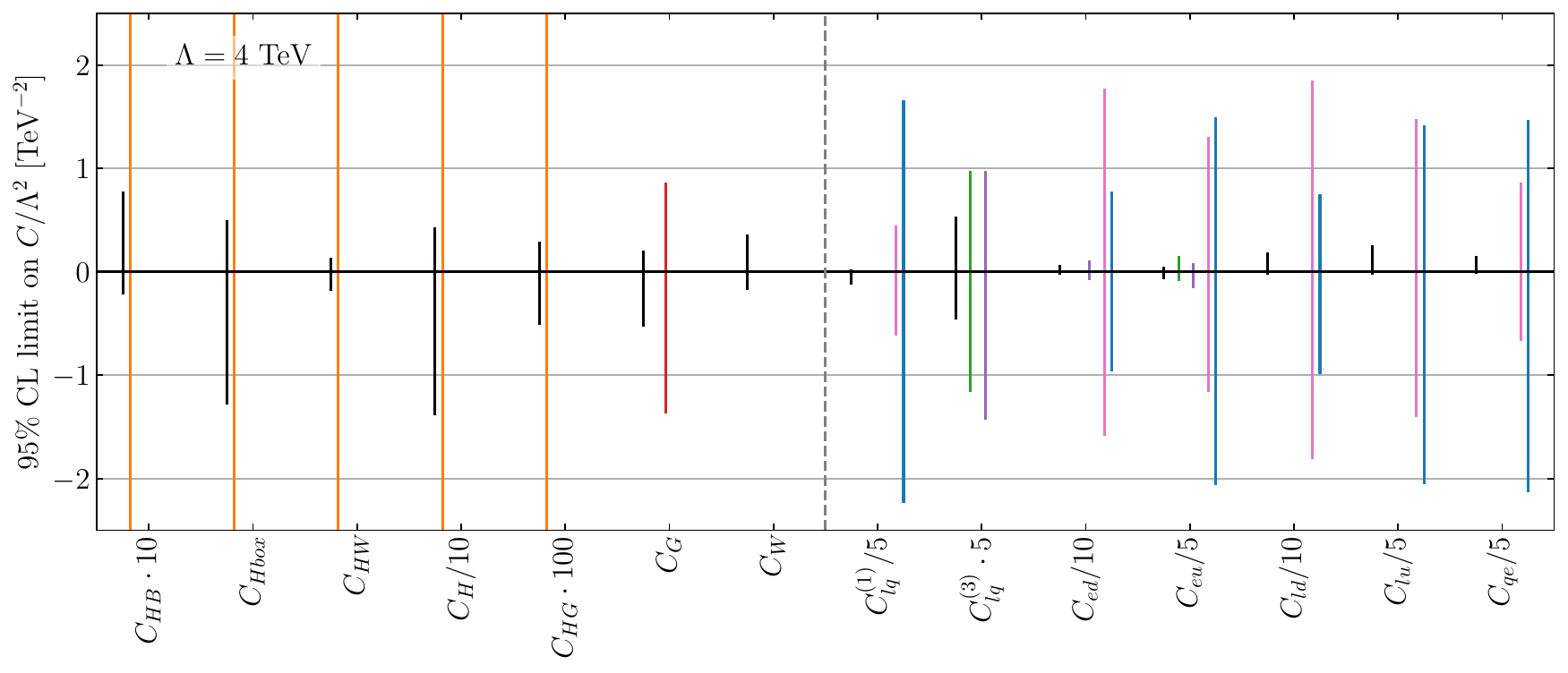}
    \includegraphics[height=6.0cm]{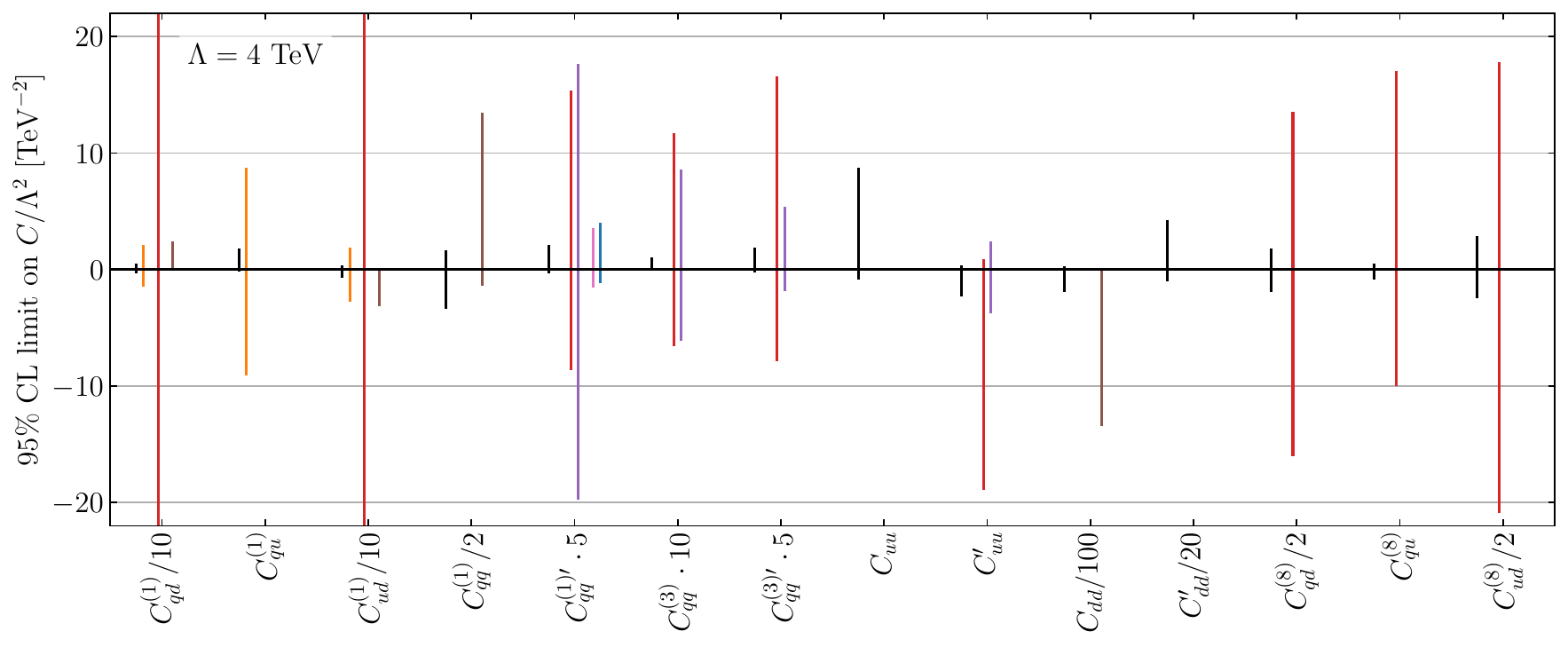}
    \caption{Global analysis including NLO predictions. Different colours correspond to analyses removing certain datasets to highlight their relevance. Note that some Wilson coefficients have been rescaled and that the y-axis range is different in the three plot panels.}
    \label{fig:NLO_fit_datasets}
\end{figure}

The datasets described in Section~\ref{sec:datasets} are sensitive to different, and often unique, directions in parameter space. The combination of different datasets hence plays a crucial role for the limits obtained in a global analysis. In this Section, we will discuss the constraining power of the individual datasets as well as their interplay. 

In Figure~\ref{fig:NLO_fit_datasets}, we once again display the results of the comprehensive global analysis encompassing all 41 Wilson coefficients. 
In addition to the global analysis including all datasets, we use different colours to represent fits with the exclusion of a single dataset to highlight the relevance of the respective set. 
To ensure a better readability of the plot, we exclusively display bounds in cases where the removal of a dataset impacts the constrained range by a minimum of 100~\%.
As expected, the ten operators contributing to EWPO at LO, displayed in the upper panel,  are dominantly constrained by EWPO and Higgs physics, corresponding to the orange and green lines, respectively. 
However, we see that the other datasets, such as PVE (pink), top (red) and dijets (brown), are also crucial for constraining some of these operators. 
For $C_{Hl}^{(1)}$, the relevance of the PVE dataset is a result of correlations with semileptonic operators and will be further explored in Section~\ref{sec:EWPO_PVE}.
For $C_{Hd}$ and $C_{Hq}^{(1)}$, correlations in EWPO with four-quark operators ($C_{Hq}^{(1)}$ with $C_{qq}^{(1)}$, $C_{uu}$ and $C_{Hd}$ with the weakly constrained $C_{dd}$) lead to relevant contributions from top and dijet observables. 

Four-lepton operators are dominantly constrained in lepton-scattering experiments and EWPO. 
By constraining $C_{ll}^\prime$, EWPO also play an important role in disentangling the correlations with $C_{ll}$ present in lepton scattering.

Operators contributing to modified Higgs couplings, shown in the central panel, do not receive important contributions from any other sector. $C_G$ and $C_W$ are dominantly constrained in top physics and $Zjj$ production (the latter is not shown as an individual dataset), respectively.

Semileptonic operators, also displayed in the central panel, are all dominantly constrained by PVE and Drell-Yan, except $C_{lq}^{(3)}$ and $C_{eu}$ which receive important contributions from the flavour sector as well as EWPO. 
We study the interplay of PVE and Drell-Yan further in Section~\ref{sec:DY_PVE}. 

\begin{figure}
    \centering
    \includegraphics[width=0.45\textwidth]{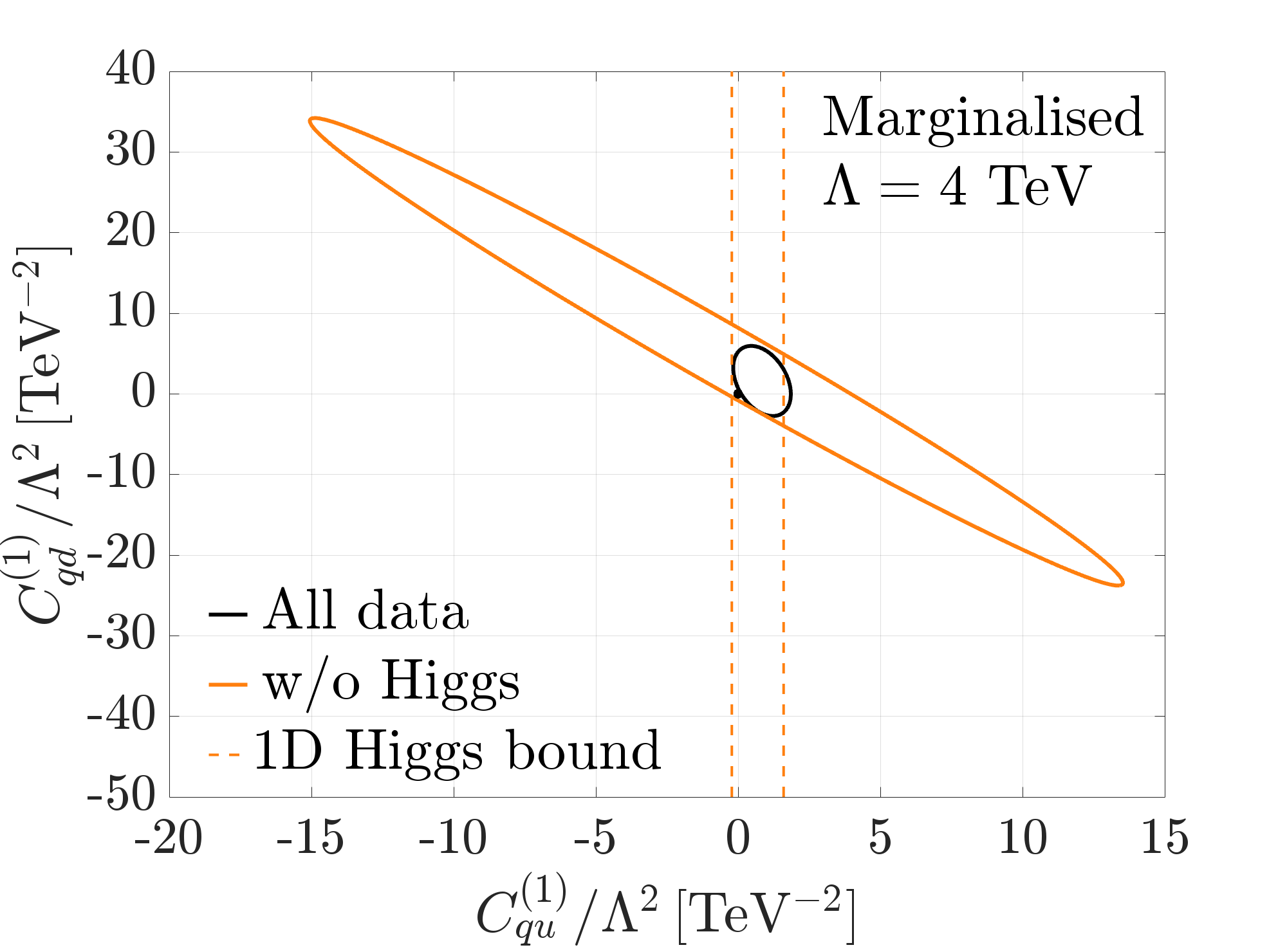}
    \includegraphics[width=0.45\textwidth]{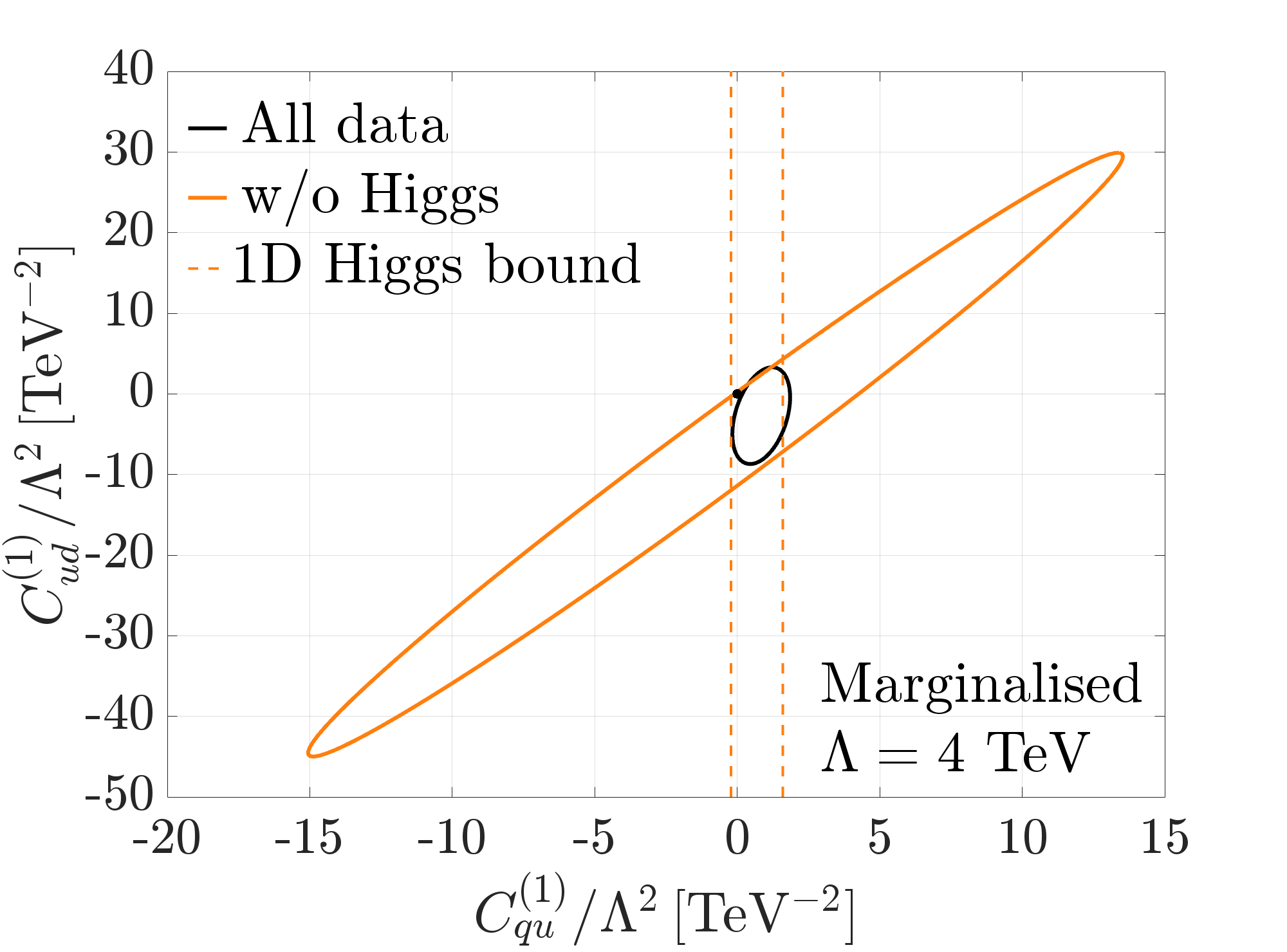}
    \caption{95\% CI limits with and without the inclusion of the Higgs dataset showing the impact of this dataset on $C_{qd}^{(1)}$ (left) and $C_{ud}^{(1)}$ (right) through lifting correlations with 
    $C_{qu}^{(1)}$.
    }
    \label{fig:Higgs_impact_Cqu1}
\end{figure}

Interestingly, four-quark operators, displayed in the lower panel, receive important constraints from all considered datasets except EWPO. For most operators the most dominant constraints result from Higgs, top, dijet and flavour data. 
Higgs physics plays an important role in setting the limits on $C_{qd}^{(1)}, \, C_{qu}^{(1)}, \, C_{ud}^{(1)}$, even though only $C_{qu}^{(1)}$ is directly constrained in Higgs physics (through NLO $t \bar{t}h$ production). 
The effect on the other two operators is the result of strong correlations with $C_{qu}^{(1)}$ as shown in Figure~\ref{fig:Higgs_impact_Cqu1}. 
The orange contours correspond to a global analysis without Higgs data and show a strong correlation between the considered four-quark operators. Limits on $C_{qu}^{(1)}$ from Higgs physics, indicated by the dashed lines, break this degeneracy. 
The combination of all datasets (black contours) therefore not only enhances the limits on $C_{qu}^{(1)}$ but also influences the correlated coefficients $C_{qd}^{(1)}$ and $C_{ud}^{(1)}$. 

For $C_{qd}^{(1)}$, $C_{ud}^{(1)}$, $C_{qq}^{(1)}$, $C_{dd}$, dijet data have a relevant impact for breaking degeneracies between operators and shifting the limits towards SM values. Without dijet data, deviations in $t\bar{t}h$ production push the limits towards non-SM values. 
For $C_{dd}^\prime$, which can only be very weakly constrained, the changes of the limits when removing the EWPO and dijet dataset are smaller than the chosen threshold of two (100\%). Therefore, no dataset-specific additional limits are shown in the plot.
Somewhat surprisingly, Drell-Yan and PVE influence the bounds on $C_{qq}^{(1) \prime}$. We further investigate this effect in Section~\ref{sec:DY_PVE}.

\subsection{EWPO and PVE datasets}
\label{sec:EWPO_PVE}
\begin{figure}[thb]
    \centering
    \includegraphics[width=0.45\textwidth]{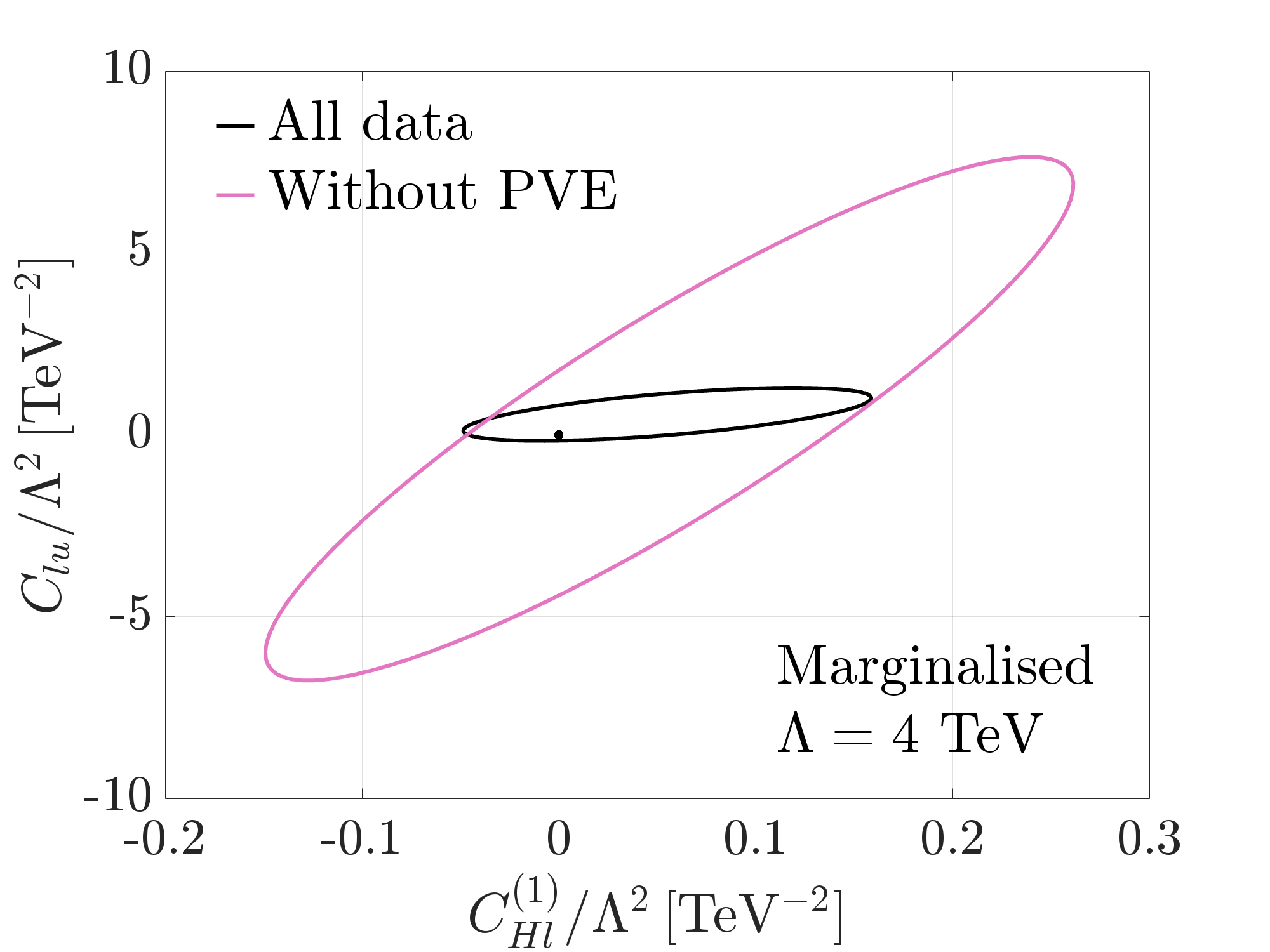}
    \includegraphics[width=0.45\textwidth]{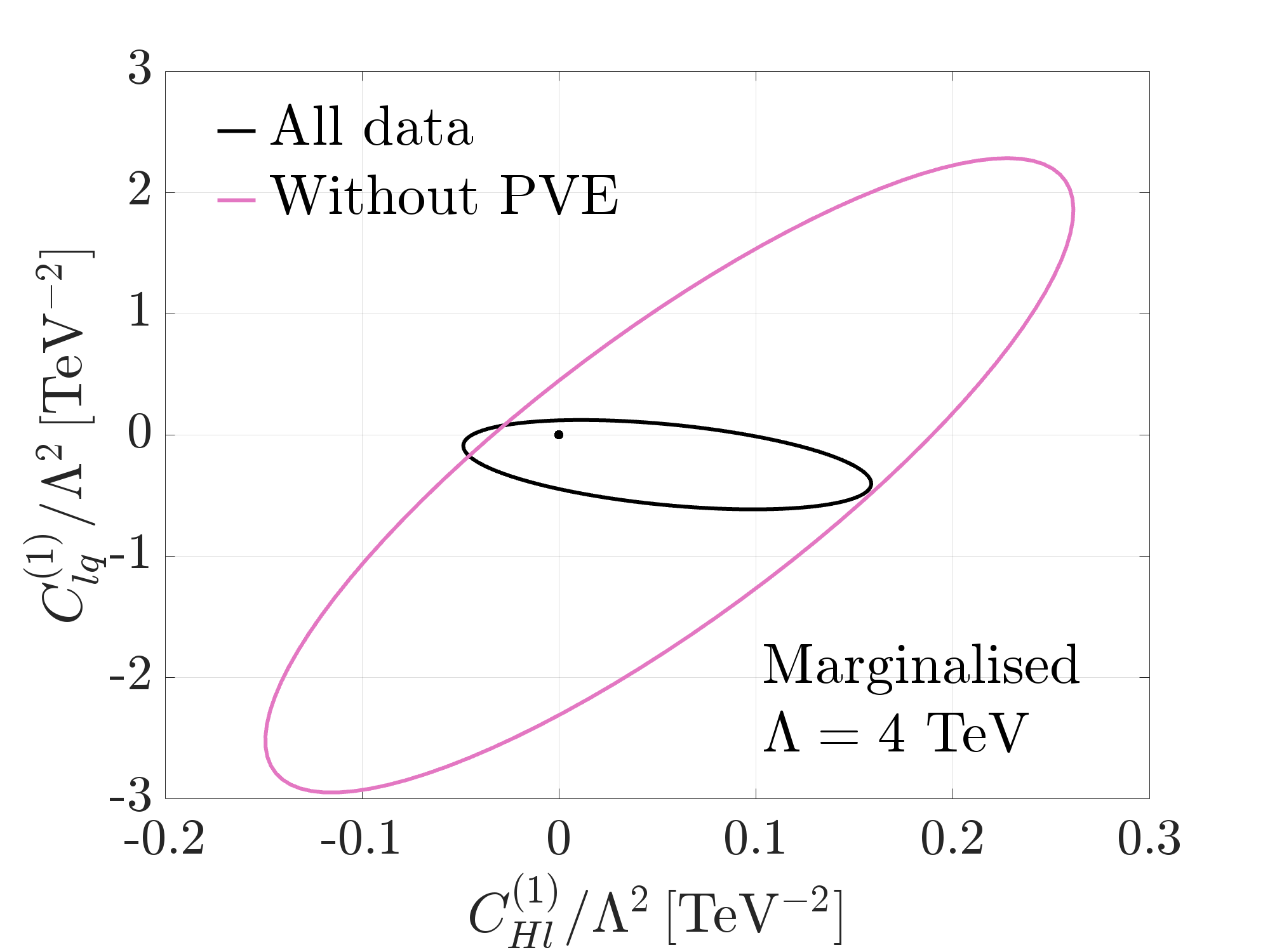}
    \caption{95\% CI limits with and without the inclusion of the PVE dataset showing the impact of this dataset on $C_{Hl}^{(1)}$ through lifting correlations with $C_{lu}$ and $C_{lq}^{(1)}$.}
    \label{fig:CHl1_PVE_impact}
\end{figure}
At LO, the operators contributing to the description of EWPO are tightly constrained. At NLO, the number of operators contributing to EWPO extends from ten to 35, hence adding a lot more freedom to the fit. 
The additional degrees of freedom need to be constrained by other datasets, otherwise the limits on the operators appearing at LO will be degraded. 
PVE plays an important role in constraining semileptonic operators and lifting degeneracies between these and the operators entering EWPO at LO. We illustrate two examples of the breaking of this degeneracy in Figure~\ref{fig:CHl1_PVE_impact}. 
The pink contours show large correlations between $C_{Hl}^{(1)}$ and $C_{lu}$ (left panel) as well as $C_{lq}^{(1)}$ (right panel) when PVE data are absent from the fit. 
By setting strong constraints on $C_{lu}$ and $C_{lq}^{(1)}$ and testing different directions in the $C_{Hl}^{(1)}$--$C_{lu}$/$C_{lq}^{(1)}$ parameter space, PVE has an important impact on the global analyses bounds of $C_{Hl}^{(1)}$, see Figure~\ref{fig:NLO_fit_datasets}. 

In~\cite{Cirigliano:2022qdm,ThomasArun:2023wbd,Cirigliano:2023nol}, it was shown that the inclusion of $\beta$-decay observables and semileptonic meson decays~\cite{Gonzalez-Alonso:2016etj,Falkowski:2020pma} can have an interesting interplay with EWPO. In appendix~\ref{app:dckm}, we estimate this effect by adding $\Delta_{\rm CKM}$ as a pseudo-observable to our fit.

\subsection{Drell-Yan, PVE and flavour datasets}
\label{sec:DY_PVE}
\begin{figure}[thb]
\centering
\includegraphics[height=5.3cm]
{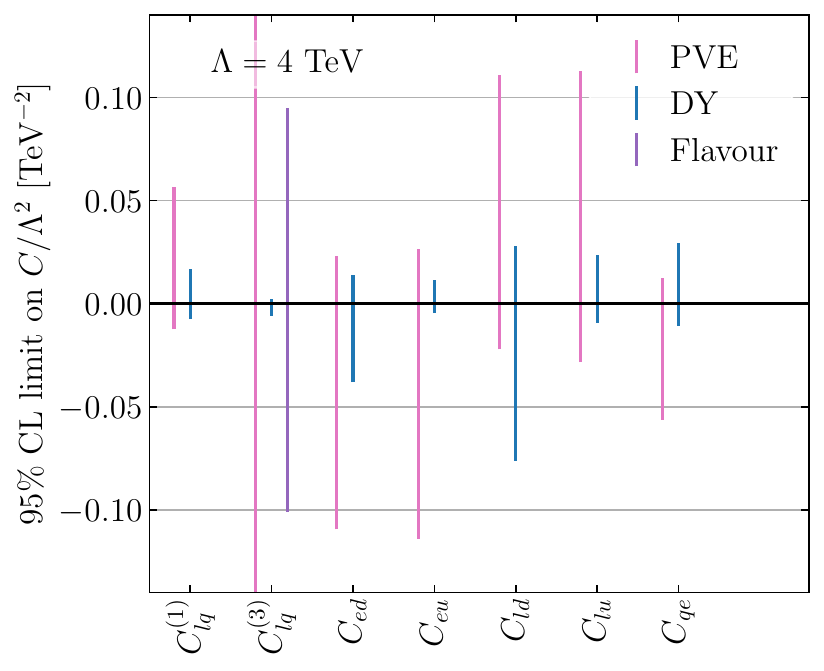}
\includegraphics[height=5.6cm]{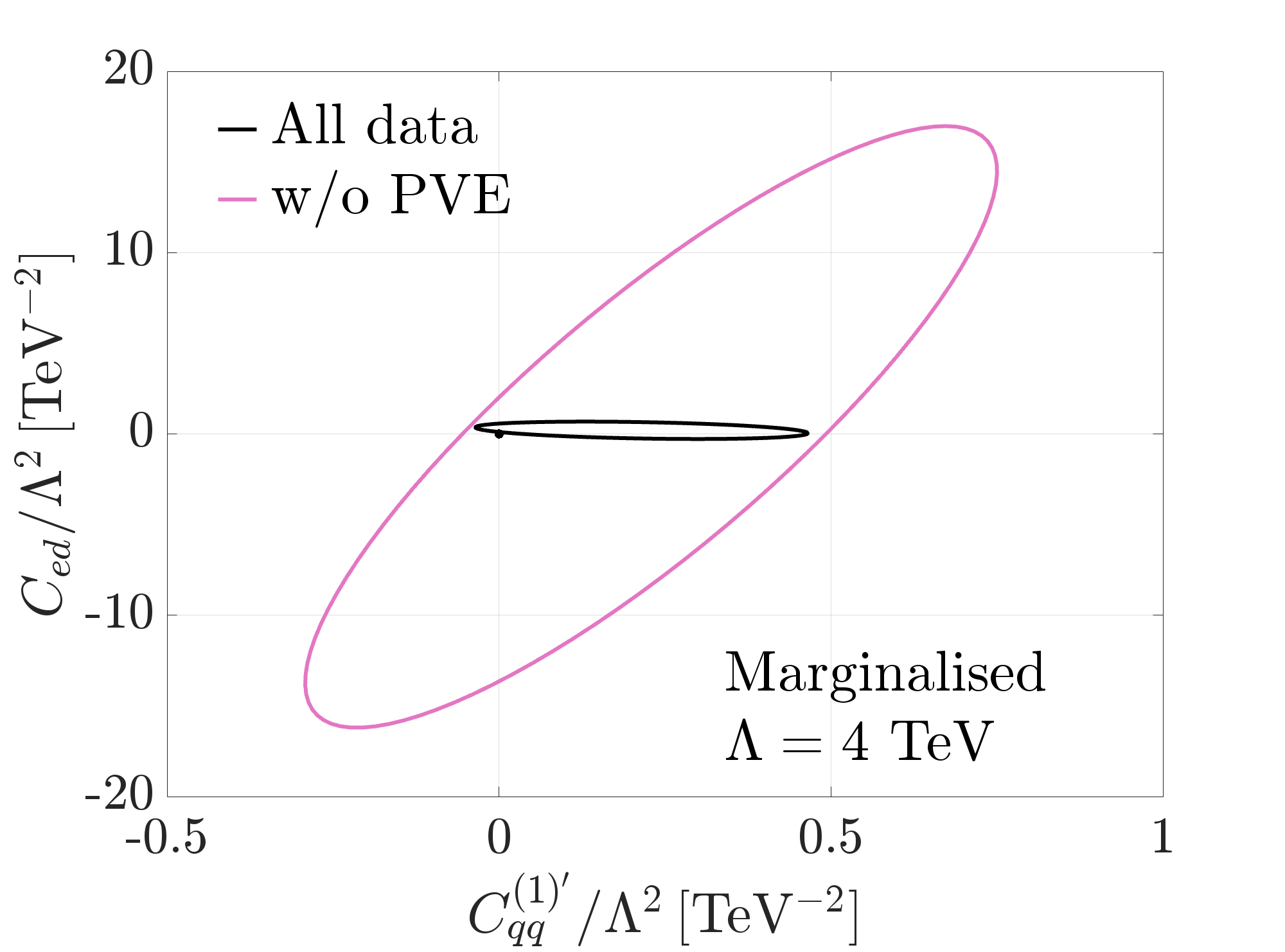}
   \caption{(Incomplete) list of single-parameter bounds at $95\%$ CL on semileptonic operators (left). Note that for the flavour dataset we suppress the limits on all operators except $C_{lq}^{(3)}$ because they are not competitive with PVE and Drell-Yan. 95\% CI limits with and without the inclusion of the PVE dataset showing the impact of this dataset on $C_{qq}^{(1)\prime}$ through lifting correlations with $C_{ed}$ (right).}
\label{fig:single_PVE_DY}
\end{figure}

\begin{figure}[H]
    \centering
    \includegraphics[width=0.45\textwidth]{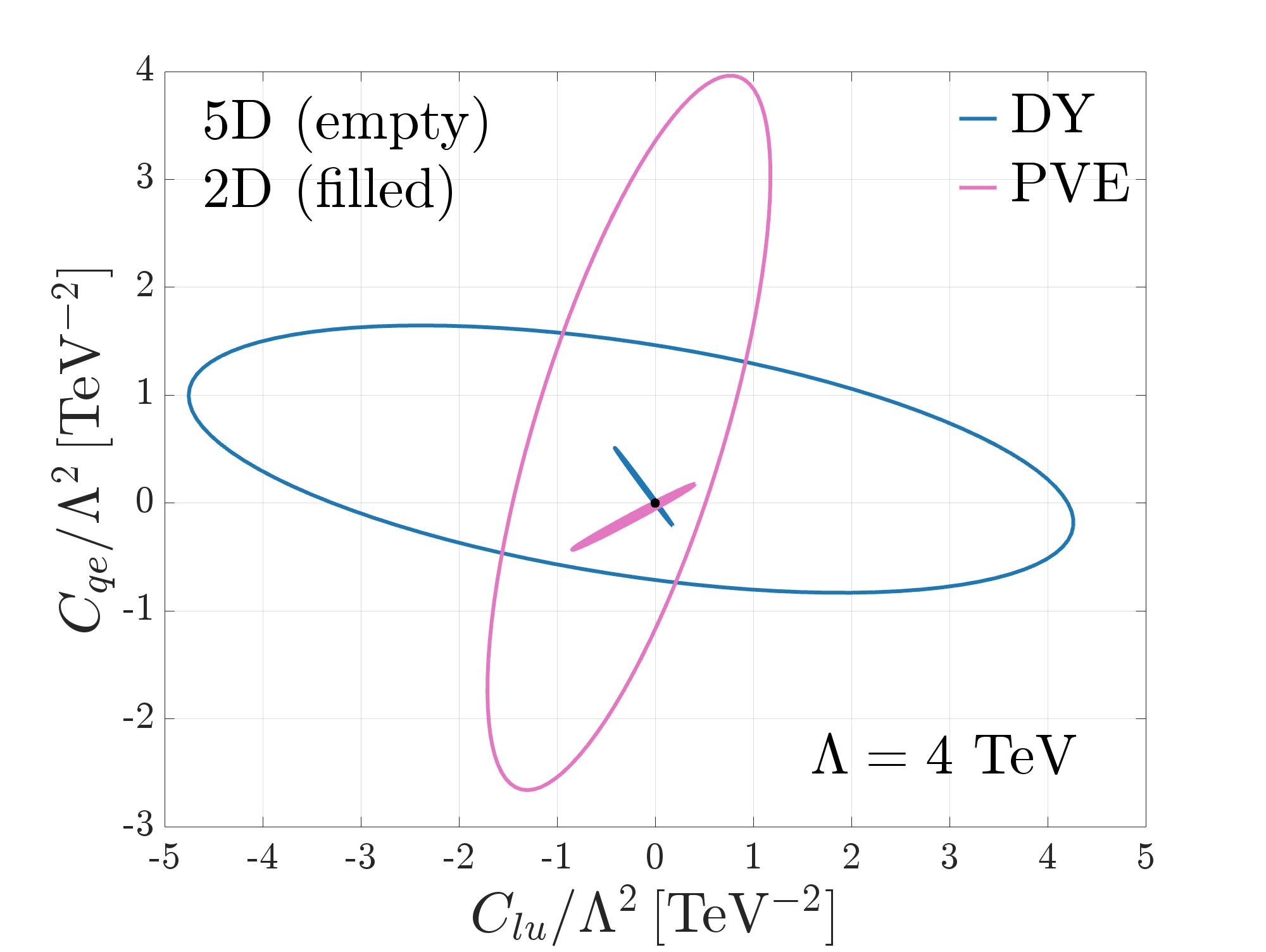}
    \includegraphics[width=0.45\textwidth]{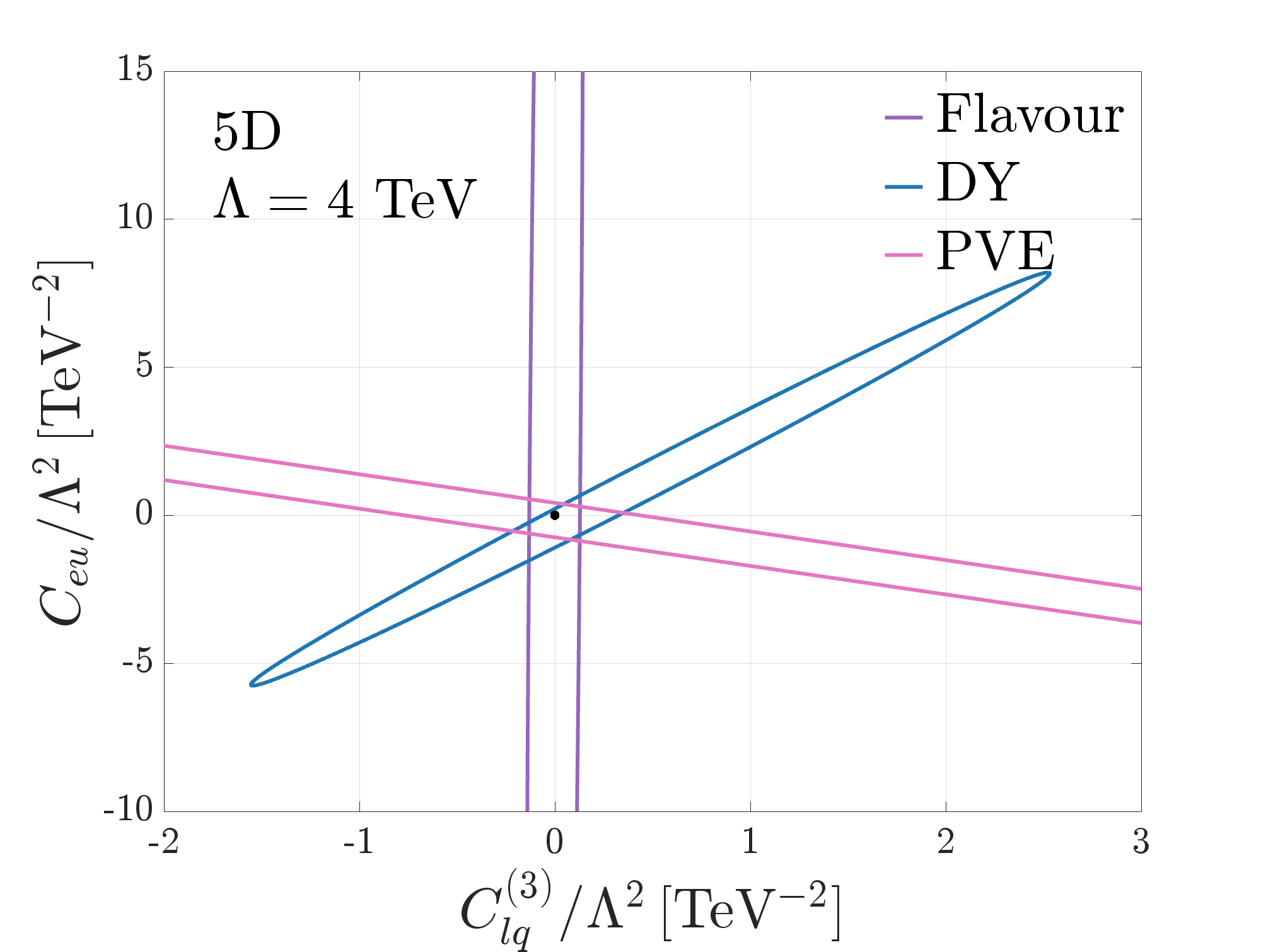}
    \caption{$95\%$ CI contours on the two-dimensional parameter space of the semileptonic operators from PVE and Drell-Yan (and flavour). The 5D analyses are marginalised over the Wilson coefficients $C_{lq}^{(1)}$, $C_{lq}^{(3)}$ and $C_{eu}$ (left) and $C_{lq}^{(1)}$, $C_{lu}$ and $C_{qe}$ (right).}
    \label{fig:PVE-DY-2d}
\end{figure}

Drell-Yan and PVE data are both sensitive to semileptonic operators
\begin{equation}
\{C_{lq}^{(1)},C_{lq}^{(3)},C_{ed},C_{eu},C_{ld},C_{lu},C_{qe}\} \, .
\label{ops:PVE-DY}
\end{equation}
We present single-parameter fit results for both datasets in Figure~\ref{fig:single_PVE_DY}. For $C_{lq}^{(3)}$, we also show limits from the flavour dataset, which are competitive for this coefficient. 
Constraints from Drell-Yan dominate over those from PVE throughout the whole parameter space. Nevertheless, limits from PVE are of a similar order of magnitude and this dataset plays an important role in the global analysis through its ability to test different directions in parameter space compared to Drell-Yan. 
To highlight the complementarity of both searches, we present 2D-limit plots for $C_{lu}$--vs--$C_{qe}$ and $C_{lq}^{(3)}$--vs--$C_{eu}$ in Figure~\ref{fig:PVE-DY-2d}. 
In a two-parameter fit, shown by the filled contours, we find that PVE (pink) and Drell-Yan (blue) probe almost orthogonal directions in parameter space for $C_{qe}$ and $C_{lu}$. 
In Drell-Yan, these two Wilson coefficients appear with the same sign as they are both dominated by up-type quark contributions, which leads to an anti-correlation of $C_{qe}$ and $C_{lu}$. Only exclusive down-type quark contributions, $C_{ed}$ and $C_{ld}$, as well as the structurally distinct $C_{lq}^{(3)}$ enter with an opposite sign for Drell-Yan. 
The PVE dataset is sensitive to axial couplings in the quark or lepton sector. As the operators corresponding to $C_{qe}$ and $C_{lu}$ involve quarks and leptons of different chiralities, all PVE observables will be sensitive to a negative relative combination of these two operators (whenever both operators appear), which results in a positive correlation of the Wilson coefficients. 
The complementarity between the PVE and Drell-Yan datasets also remains when marginalising over additional semileptonic Wilson coefficients. This is shown by the unfilled contours in the left panel of Figure~\ref{fig:PVE-DY-2d}. 
Comparing Figure~\ref{fig:NLO_fit_datasets}, we find that PVE is particularly relevant for the semileptonic operators specifically involving down-type quarks since the corresponding measurements do not experience the same parton distribution function (PDF) suppression as Drell-Yan. 

In the right panel of Figure~\ref{fig:PVE-DY-2d}, we show the two-parameter limits on $C_{lq}^{(3)}$--vs--$C_{eu}$ when marginalising over $C_{lq}^{(1)}$, $C_{lu}$ and $C_{qe}$. Again, Drell-Yan and PVE probe complementary directions in parameter space. In addition, flavour plays an important role by setting constraints on $C_{lq}^{(3)}$. 

Even when combining all datasets, the limits on almost all semileptonic operators are weakened by a more than a factor $20$ in the global analysis, see Figure~\ref{fig:ratio_global_1D}.
The Drell-Yan dataset delivers strong individual constraints on semileptonic operators. However, different bins and final states of the distributions only have a marginal impact on the relative importance of different Wilson coefficients of semileptonic operators. Therefore, this dataset is particularly affected by correlations.
Therefore, in order to further improve bounds on semileptonic operators, it is relevant to add observables constraining different combinations of the corresponding Wilson coefficients. 

The complementarity between low-energy Parity-Violating Electron Scattering (PVES) and LHC Run~I Drell-Yan data in constraining semileptonic operators has already been studied in~\cite{Boughezal:2021kla}. While this previous work is based on LHC Run~I data up to $1.5$~TeV, our analysis includes LHC Run~II data up to an invariant-mass of $3$~TeV.

In addition to setting limits on semileptonic operators, we have observed in  Figure~\ref{fig:NLO_fit_datasets} that the PVE and Drell-Yan datasets also affect bounds on the four-quark operator $C_{qq}^{(1)\prime}$ even though they are not directly sensitive to this coefficient. 
To explain this feature, we report in the right panel of Figure~\ref{fig:single_PVE_DY} the 2D-limits on $C_{qq}^{(1)\prime}$--vs--$C_{ed}$. 
In a global fit without PVE data, these two operators are correlated as shown by the pink contour. 
PVE sets a stringent constraint on $C_{ed}$ which in turn improves the bounds on $C_{qq}^{(1)\prime}$ significantly, see the black contour.

\subsection{Top, flavour and dijet datasets}
\begin{figure}[thb]
\centering
\includegraphics[width=0.85\textwidth]{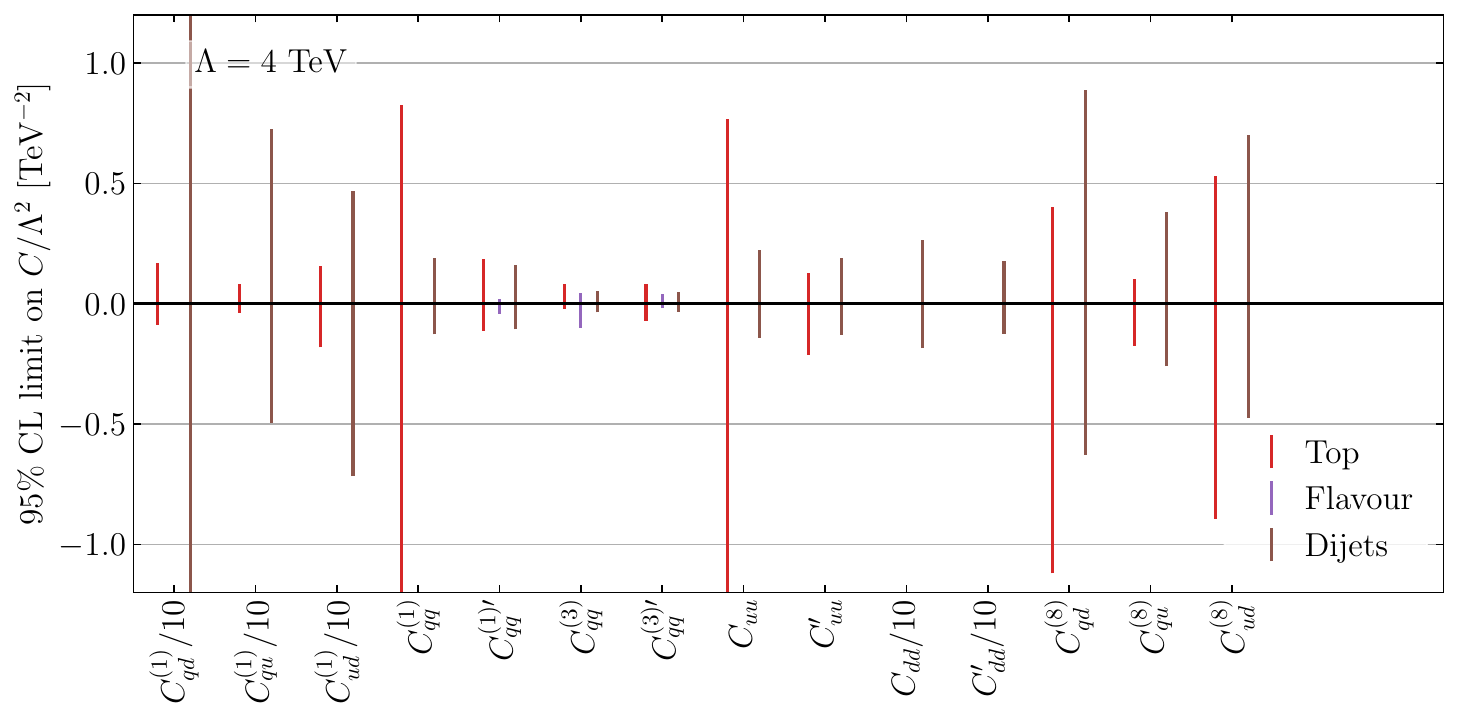}
   \caption{Single-operator bounds at $95\%$ CL on the Wilson coefficients of the four-quark operators from different datasets. }
\label{fig:single_top_flav}
\end{figure}

Top, flavour and dijet data overlap in their sensitivity to four-quark operators. 
In the following, we will discuss in how far the bounds from these sectors are complementary in a global fit, focusing in particular on the interplay of top and flavour data. 

\begin{figure}[thb]
    \centering
    \includegraphics[width=.45\textwidth]{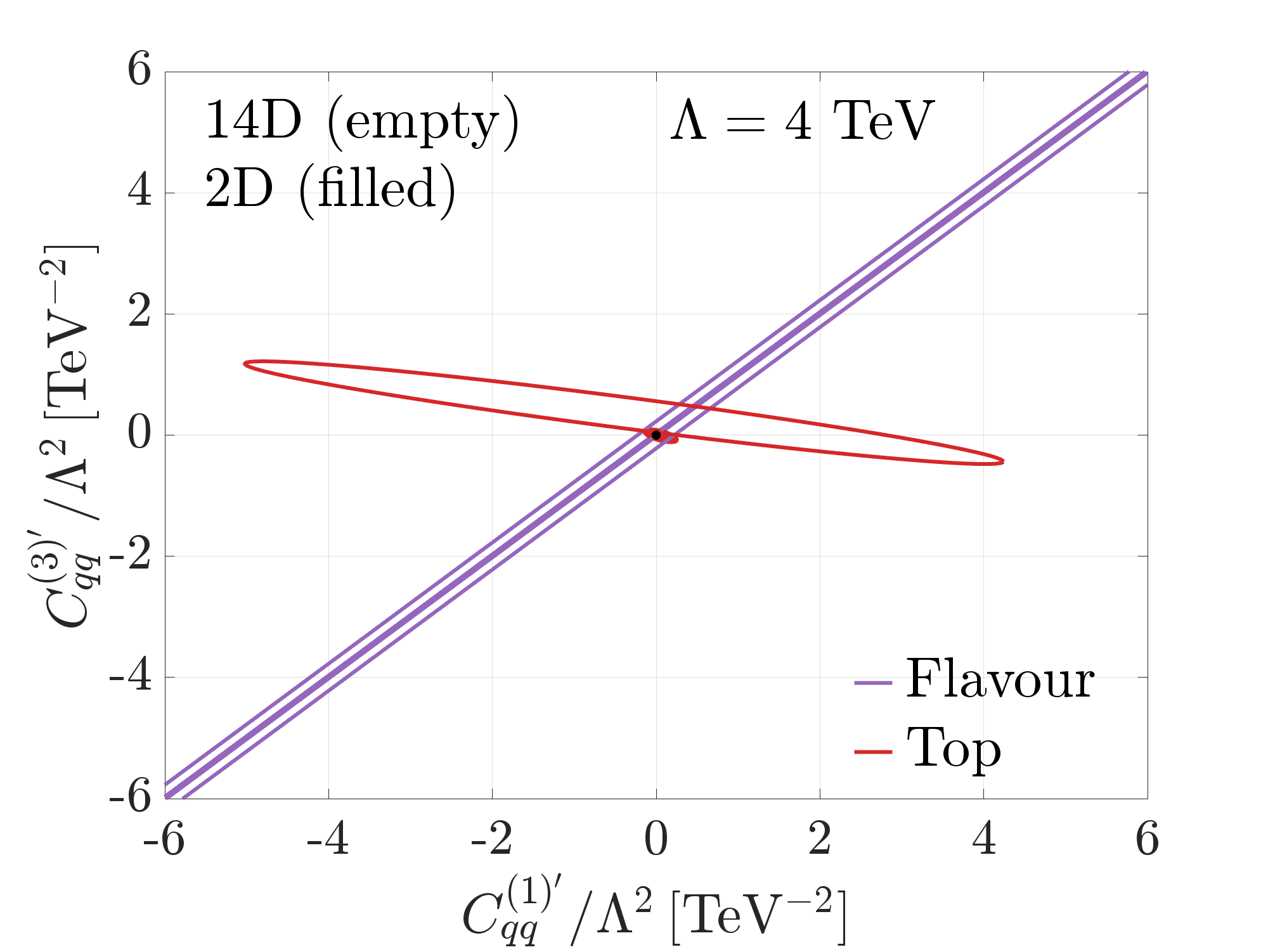}
    \caption{$95\%$ CI contours of $C_{qq}^{(1)\prime}$ vs $C_{qq}^{(3)\prime}$ from flavour and top data. See text for details. }
    \label{fig:top-flav-2d}
\end{figure}

We compare the one-parameter bounds on the relevant Wilson coefficients in Figure~\ref{fig:single_top_flav}. 
At the single-parameter level, top (red) and dijet (brown) data are equally relevant for constraining four-quark operators. However, at the level of a global fit, limits from top physics typically dominate and dijet data is mainly relevant to set limits on operators which do not contribute to top physics such as those involving only down-type quarks, see Figure~\ref{fig:NLO_fit_datasets}. 
The flavour dataset (purple) has a crucial impact on the bounds on $C_{qq}^{(1)\prime}$ and $C_{qq}^{(3)\prime}$ even in our $U(3)^5$ symmetric scenario, setting the strongest one-parameter bounds on these operators and having a strong effect in the global analysis as well. 
For these two parameters, flavour data can only constrain the direction $C_{qq}^{(1)\prime} - C_{qq}^{(3)\prime}$, which contributes at tree-level to $C_1$ and at one-loop to $C_9$ and $C_{10}$, see appendix~\ref{app:LEFT_Hamiltonians} for the definitions of these LEFT coefficients.
This can be explained by the fact that only structures with one up-type and one down-type quark bilinear $(\bar{u}\gamma_\mu u)_L (\bar{d} \gamma^\mu d)_L$ with colour indices contracted within the brackets contribute to flavour observables, for which $C_{qq}^{(1)\prime}$ and $C_{qq}^{(3)\prime}$ have exactly opposite signs. 
On the other hand, $t\bar{t}$ production receives contributions from both the above structure and from its equivalent with two up-type quark bilinears. 
The latter, to which $C_{qq}^{(1)\prime}$ and $C_{qq}^{(3)\prime}$ both contribute with the same sign, dominates due to the larger up-type PDFs. 
As a result, $C_{qq}^{(1)\prime}$ and $C_{qq}^{(3)\prime}$ have a positive relative sign in all considered $t\Bar{t}$ observables. 
We show the complementary behaviour of top and flavour data in Figure~\ref{fig:top-flav-2d}. The filled contours correspond to fitting $C_{qq}^{(1)\prime}$ and $C_{qq}^{(3)\prime}$ only, whereas the empty contours have been obtained in a 14-parameter fit marginalising over the remaining twelve four-quark operators.
We can see that while the limits are degraded in a global fit, as expected, the complementarity of the flavour and top datasets persists. 
Therefore, both datasets are essential in order to perform a global analysis including these two operators.
See also~\cite{Bruggisser:2022rhb} on resolving the flavour structure of the four-quark operators $Q_{qq}^{(1)}$ and $Q_{qq}^{(3)}$  including higher-order terms of the MFV expansion using similar datasets\footnote{Our parameters $C_{qq}^{(x)\prime}$  correspond to $(\widetilde{aa})^{(x)}$ in their notation.}.

In our work, the flavour dataset mostly affects bounds on the operators of the four-quark sector. When going beyond our minimal MFV hypothesis, flavour data has been shown to play an important role for constraining higher-order terms in the MFV expansion of semileptonic and gauge-fermion operators~\cite{Bruggisser:2021duo,Grunwald:2023nli,Aoude:2020dwv}.

\section{Conclusions and outlook}
\label{sec:conclusions}

We conducted a global analysis of all 41 operators of the minimal MFV SMEFT, i.e.\ assuming an exact $U(3)^5$ symmetry at the high scale. 
The selection of our Wilson-coefficient set is entirely guided by theory and no flat directions remain in the fit when considering EWPO, Higgs, top, flavour, dijet, PVE, Drell-Yan and lepton scattering datasets. 
In an analysis based solely on LO SMEFT predictions, five four-quark operators involving right-handed fermion fields remain weakly constrained in our global fit.  However, their presence in the fit does not significantly impact the limits on the remaining operators, highlighting that there is only minimal crosstalk between both Wilson-coefficient sets.

Including partial NLO SMEFT predictions enables us to test the four-quark operators which only contribute weakly to our observables at LO. 
With partial NLO SMEFT corrections, only the Wilson coefficients $C_{dd}$ and $C_{dd}^\prime$ exhibit bounds on $C/\Lambda^2$ weaker than $10/\text{TeV}^2$. 
These two Wilson coefficients are highly anti-correlated.
For another ten operators, the bounds are weaker than $1/\text{TeV}^2$, the remaining 29 operators can be constrained to $|C|/\Lambda^2 < 1/\text{TeV}^2$.
All 41 Wilson coefficients remain consistent with the SM within~$2\sigma$ in the LO and NLO analyses. Six (nine) Wilson coefficients exhibit deviations exceeding $1.5\sigma$ in the LO (NLO) fit. 

In principle, switching to NLO SMEFT predictions introduces additional degeneracies between operators as a single observable is a function of a larger set of Wilson coefficients at NLO. For the EWPO dataset, the number of relevant flavour-symmetric Wilson coefficients increases from ten at LO to 35 at NLO. 
Nevertheless, for most Wilson-coefficient bounds the impact of going to NLO precision in the EWPO dataset is weak. This shows that constraints on the operators first appearing at NLO in EWPO from other sectors are strong enough to render the effect of the additional degeneracies subdominant. 
Of the operators constrained at LO in EWPO, only the limit on $C_{Hq}^{(1)}$ is impacted significantly by correlations with four-quark operators induced at NLO and increases by a factor 2.5. 

The limits on semileptonic operators strongly increase in the global fit compared to single-parameter fits. This is due to Drell-Yan providing very strong single-parameter limits, while only very poorly disentangling the effects of different semileptonic operators, as these typically experience the same dependence on the considered energy and final-state flavour. 
The combination of Drell-Yan with PVE data, which provides weaker limits at a single-parameter fit level, is highly relevant to constrain semileptonic operators in a global analysis, as it allows to break the degeneracies present in the Drell-Yan dataset. 
In the bosonic sector, the largest limit increases in a global fit  compared to a single-parameter fit occur for $C_{HWB}$, $C_{HD}$, $C_{HB}$ and $C_{HW}$. 

Limits on four-quark operators, particularly those involving down-type quarks, profit from the inclusion of dijet data. To avoid large quadratic Wilson-coefficient contributions and therefore a potential violation of the EFT validity, we recast an ATLAS dijet+photon search~\cite{ATLAS:2019itm}, which probes the low-$m_{jj}$ invariant-mass range. SMEFT predictions for this differential distribution are included as an ancillary file with the arXiv submission. 
Data from flavour as well as top physics test complementary directions in the four-quark operator space. Therefore, their combination is essential to fully constrain the NP parameter space. 

In the future, it would be interesting to consider all SMEFT observables at NLO precision  and include renormalisation group running effects beyond the flavour sector~\cite{Aoude:2022aro}.

Our results show the necessity of combining numerous datasets at different scales for global SMEFT analyses. Datasets which seem subdominant at the single-parameter fit level can play a significant role in disentangling directions in the Wilson coefficient parameter space.

\section*{Acknowledgements}
We thank Luca Mantani and Ken Mimasu for their invaluable assistance in comparing the SMEFit and fitmaker theory predictions and the authors of~\cite{Kassabov:2023hbm} for providing additional NLO SMEFT predictions.  All three authors are supported by  the  Cluster  of  Excellence  ``Precision  Physics,  Fundamental
Interactions, and Structure of Matter" (PRISMA$^+$ EXC 2118/1) funded by the German Research Foundation (DFG) within the German Excellence Strategy (Project ID 390831469). 
A.B.~also gratefully acknowledges support from the Alexander-von-Humboldt foundation as a Feodor Lynen Fellow during large parts of this project.
T.H. also thanks the CERN theory group for its hospitality during his regular visits to CERN where part of the work was done.


\appendix
\newpage

\section{$\Delta_{\rm CKM}$ as an additional  low-energy observable}
\label{app:dckm}

\begin{figure}[t]
    \centering
    \includegraphics[width=0.65\textwidth]{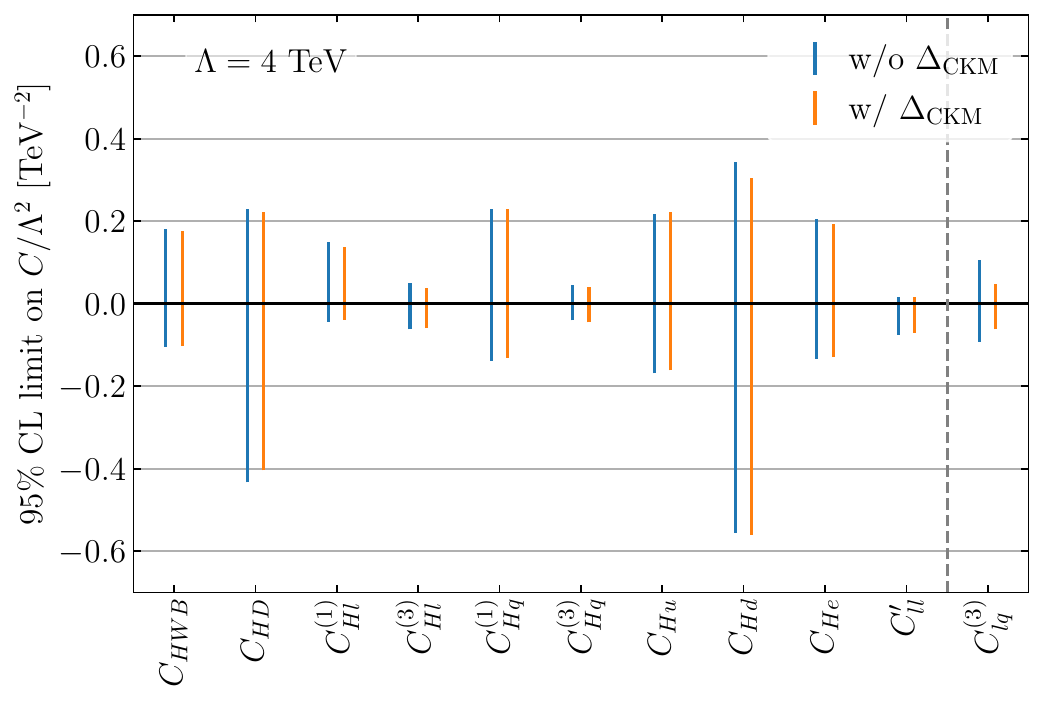}
    \caption{Global analysis of all 41 Wilson coefficients with and without the inclusion of $\Delta_{\rm CKM}$ as an observable. Only the ten Wilson coefficients contributing to EWPO at LO  and  $C_{lq}^{(3)}$ are shown.}
    \label{fig:global_dckm}
\end{figure}
\enlargethispage{5mm}

Observables from $\beta$-decay  and semileptonic meson decays~\cite{Gonzalez-Alonso:2016etj,Falkowski:2020pma}
have been shown to have interesting interplay with EWPO~\cite{Cirigliano:2022qdm,ThomasArun:2023wbd,Cirigliano:2023nol}.
In this section, we estimate the effect of including these measurements by adding a single pseudo-observable to our fit. 
The unitarity of the CKM matrix experiences a $2\sigma$ deviation from zero~\cite{ParticleDataGroup:2022pth}
\begin{equation}
    \Delta_{\rm CKM} = |V_{ud}|^2 + |V_{us}|^2 -1  = - 0.0015 \pm 0.0007 \, ,
\end{equation}
where $V_{ud}$ is dominantly constrained by superallowed nuclear $\beta$ decays and $V_{us}$ receives the strongest constraints from semileptonic Kaon decays. 
In the SMEFT, the measurements of $V_{ij}$ are influenced by modifications of the $W$-boson couplings as well as four-fermion operator contributions 
\begin{equation}
    \Delta_{\rm CKM} = 2 \,   \frac{v^2}{\Lambda^2} \left( C_{Hq}^{(3)} - C_{Hl}^{(3)} + C_{ll}^\prime - C_{lq}^{(3)} \right) \, .
\end{equation}
Including $\Delta_{\rm CKM}$ as a pseudo-observable and performing a fit, we find only a mild influence on most operators in our 41-parameter fit. The corresponding limits are shown in Figure~\ref{fig:global_dckm}.
Only the limits on $C_{lq}^{(3)}$ and $C_{Hl}^{(3)}$ are influenced by more than $10\%$ and  decrease by factors of $0.55$ and $0.86$, respectively. 
Decreasing the parameter set to the eleven coefficients explicitly shown in Figure~\ref{fig:global_dckm} and removing the Higgs dataset, we find a stronger impact of the observable $\Delta_{\rm CKM}$ with four of the coefficients contributing to EWPO at LO changing by more than $10\%$, thus confirming the findings of~\cite{Cirigliano:2022qdm,Cirigliano:2023nol}.  $C_{lq}^{(3)}$ is tightly constrained by Drell-Yan data in the eleven-parameter fit and therefore changes by less than $5\%$ when including $\Delta_{\rm CKM}$ as a pseudo-observable in this case.

It would be interesting to properly include observables from $\beta$-decay  and semileptonic meson decays~\cite{Gonzalez-Alonso:2016etj,Falkowski:2020pma} in the global analysis, even though we expect the impact to be diminished with respect to low-dimensional Wilson-coefficient analyses~\cite{Cirigliano:2022qdm,Cirigliano:2023nol}.

\newpage
\section{Flavour symmetric and CP even operators}
\label{app:operators}

\begin{table}[h]
\begin{center}
\small
\begin{minipage}[t]{4.4cm}
\renewcommand{\arraystretch}{1.5}
\begin{tabular}[t]{c|c}
\multicolumn{2}{c}{$1:X^3$} \\
\hline
$Q_G$                & $f^{ABC} G_\mu^{A\nu} G_\nu^{B\rho} G_\rho^{C\mu} $ \\
$Q_W$                & $\epsilon^{IJK} W_\mu^{I\nu} W_\nu^{J\rho} W_\rho^{K\mu}$ \\ 
\end{tabular}
\end{minipage}
\begin{minipage}[t]{2.6cm}
\renewcommand{\arraystretch}{1.5}
\begin{tabular}[t]{c|c}
\multicolumn{2}{c}{$2:H^6$} \\
\hline
$Q_H$       & $(H^\dag H)^3$ 
\end{tabular}
\end{minipage}
\begin{minipage}[t]{5.1cm}
\renewcommand{\arraystretch}{1.5}
\begin{tabular}[t]{c|c}
\multicolumn{2}{c}{$3:H^4 D^2$} \\
\hline
$Q_{H\Box}$ & $(H^\dag H)\Box(H^\dag H)$ \\
$Q_{H D}$   & $\ \left(H^\dag D_\mu H\right)^* \left(H^\dag D_\mu H\right)$ 
\end{tabular}
\end{minipage}

\vspace{0.25cm}

\begin{minipage}[t]{4.6cm}
\renewcommand{\arraystretch}{1.5}
\begin{tabular}[t]{c|c}
\multicolumn{2}{c}{$4:X^2H^2$} \\
\hline
$Q_{H G}$     & $H^\dag H\, G^A_{\mu\nu} G^{A\mu\nu}$ \\
$Q_{H W}$     & $H^\dag H\, W^I_{\mu\nu} W^{I\mu\nu}$ \\
$Q_{H B}$     & $ H^\dag H\, B_{\mu\nu} B^{\mu\nu}$ \\
$Q_{H WB}$     & $ H^\dag \tau^I H\, W^I_{\mu\nu} B^{\mu\nu}$ \\
\end{tabular}
\end{minipage}
\begin{minipage}[t]{5.2cm}
\renewcommand{\arraystretch}{1.5}
\begin{tabular}[t]{c|c}
\multicolumn{2}{c}{$7:\psi^2H^2 D$} \\
\hline
$Q_{H l}^{(1)}$      & $(H^\dag i\overleftrightarrow{D}_\mu H)(\bar l_p \gamma^\mu l_p)$\\
$Q_{H l}^{(3)}$      & $(H^\dag i\overleftrightarrow{D}^I_\mu H)(\bar l_p \tau^I \gamma^\mu l_p)$\\
$Q_{H e}$            & $(H^\dag i\overleftrightarrow{D}_\mu H)(\bar e_p \gamma^\mu e_p)$\\
$Q_{H q}^{(1)}$      & $(H^\dag i\overleftrightarrow{D}_\mu H)(\bar q_p \gamma^\mu q_p)$\\
$Q_{H q}^{(3)}$      & $(H^\dag i\overleftrightarrow{D}^I_\mu H)(\bar q_p \tau^I \gamma^\mu q_p)$\\
$Q_{H u}$            & $(H^\dag i\overleftrightarrow{D}_\mu H)(\bar u_p \gamma^\mu u_p)$\\
$Q_{H d}$            & $(H^\dag i\overleftrightarrow{D}_\mu H)(\bar d_p \gamma^\mu d_p)$\\
\end{tabular}
\end{minipage}

\vspace{0.25cm}

\begin{minipage}[t]{4.75cm}
\renewcommand{\arraystretch}{1.5}
\begin{tabular}[t]{c|c}
\multicolumn{2}{c}{$8:(\bar LL)(\bar LL)$} \\
\hline
$Q_{\ell \ell}$        & $(\bar l_p \gamma_\mu l_p)(\bar l_s \gamma^\mu l_s)$ \\
$Q_{\ell \ell}^\prime$        & $(\bar l_p \gamma_\mu l_s)(\bar l_s \gamma^\mu l_p)$ \\
$Q_{qq}^{(1)}$  & $(\bar q_p \gamma_\mu q_p)(\bar q_s \gamma^\mu q_s)$ \\
$Q_{qq}^{(3)}$  & $(\bar q_p \gamma_\mu \tau^I q_p)(\bar q_s \gamma^\mu \tau^I q_s)$ \\
$Q_{qq}^{(1)\prime}$  & $(\bar q_p \gamma_\mu q_p)(\bar q_s \gamma^\mu q_s)$ \\
$Q_{qq}^{(3)\prime}$  & $(\bar q_p \gamma_\mu \tau^I q_s)(\bar q_s \gamma^\mu \tau^I q_p)$ \\
$Q_{\ell q}^{(1)}$                & $(\bar l_p \gamma_\mu l_p)(\bar q_s \gamma^\mu q_s)$ \\
$Q_{\ell q}^{(3)}$                & $(\bar l_p \gamma_\mu \tau^I l_p)(\bar q_s \gamma^\mu \tau^I q_s)$ 
\end{tabular}
\end{minipage}
\begin{minipage}[t]{5.25cm}
\renewcommand{\arraystretch}{1.5}
\begin{tabular}[t]{c|c}
\multicolumn{2}{c}{$8:(\bar RR)(\bar RR)$} \\
\hline
$Q_{ee}$               & $(\bar e_p \gamma_\mu e_p)(\bar e_s \gamma^\mu e_s)$ \\
$Q_{uu}$        & $(\bar u_p \gamma_\mu u_p)(\bar u_s \gamma^\mu u_s)$ \\
$Q_{uu}^\prime$        & $(\bar u_p \gamma_\mu u_s)(\bar u_s \gamma^\mu u_p)$ \\
$Q_{dd}$        & $(\bar d_p \gamma_\mu d_p)(\bar d_s \gamma^\mu d_s)$ \\
$Q_{dd}^\prime$        & $(\bar d_p \gamma_\mu d_s)(\bar d_s \gamma^\mu d_p)$ \\
$Q_{eu}$                      & $(\bar e_p \gamma_\mu e_p)(\bar u_s \gamma^\mu u_s)$ \\
$Q_{ed}$                      & $(\bar e_p \gamma_\mu e_p)(\bar d_s\gamma^\mu d_s)$ \\
$Q_{ud}^{(1)}$                & $(\bar u_p \gamma_\mu u_p)(\bar d_s \gamma^\mu d_s)$ \\
$Q_{ud}^{(8)}$                & $(\bar u_p \gamma_\mu T^A u_p)(\bar d_s \gamma^\mu T^A d_s)$ \\
\end{tabular}
\end{minipage}
\begin{minipage}[t]{4.75cm}
\renewcommand{\arraystretch}{1.5}
\begin{tabular}[t]{c|c}
\multicolumn{2}{c}{$8:(\bar LL)(\bar RR)$} \\
\hline
$Q_{le}$               & $(\bar l_p \gamma_\mu l_p)(\bar e_s \gamma^\mu e_s)$ \\
$Q_{lu}$               & $(\bar l_p \gamma_\mu l_p)(\bar u_s \gamma^\mu u_s)$ \\
$Q_{ld}$               & $(\bar l_p \gamma_\mu l_p)(\bar d_s \gamma^\mu d_s)$ \\
$Q_{qe}$               & $(\bar q_p \gamma_\mu q_p)(\bar e_s \gamma^\mu e_s)$ \\
$Q_{qu}^{(1)}$         & $(\bar q_p \gamma_\mu q_p)(\bar u_s \gamma^\mu u_s)$ \\ 
$Q_{qu}^{(8)}$         & $(\bar q_p \gamma_\mu T^A q_p)(\bar u_s \gamma^\mu T^A u_s)$ \\ 
$Q_{qd}^{(1)}$ & $(\bar q_p \gamma_\mu q_p)(\bar d_s \gamma^\mu d_s)$ \\
$Q_{qd}^{(8)}$ & $(\bar q_p \gamma_\mu T^A q_p)(\bar d_s \gamma^\mu T^A d_s)$\\
\end{tabular}
\end{minipage}
\end{center}
\caption{\label{tab:basis}
Flavour symmetric and CP even dimension-six SMEFT operators in the Warsaw basis. }
\end{table}

\newpage
\section{Observables}
\label{app:observables}
We list the observables included in our analysis in Tables~\ref{tab:obset}-\ref{tab:obset_PVE_flavour}.
Table~\ref{tab:obset} lists observables from Higgs physics and $Zjj$~production, Tables~\ref{tab:obset_top} and~\ref{tab:obset_top2_DY_dijet} list observables from the top sector as well as Drell-Yan and dijet+photon data, and Table~\ref{tab:obset_PVE_flavour} lists observables from PVE, lepton scattering and flavour.  

\begin{table}[h]
	\centering
	\renewcommand{\arraystretch}{2.0}
    \caption{Higgs and electroweak observables included in the fit.}
	\begin{adjustbox}{width=0.9\textwidth}
		\label{tab:obset}
		\begin{tabular}{|c|c|c|c|}
			\hline
			\multicolumn{2}{|c|}{Observables} & no. of measurements	 &  References \\
			\hline
			\multicolumn{2}{|c|}{\bf{Higgs Data}} & 159 & \multirow{1}{*}{}  \\ 
			\cline{1-3}
			\multirow{3}{*}{7 and 8 TeV } & ATLAS \& CMS combination  & \multirow{1}{*}{20} & \multirow{1}{*}{Table~8 of Ref.~\cite{Khachatryan:2016vau}} \\
			\cline{2-4}
			\multirow{3}{*}{Run-I data }& ATLAS \& CMS combination $\mu( h \to \mu \mu)$ &  \multirow{1}{*}{1} &  \multirow{1}{*}{Table~13 of Ref.~\cite{Khachatryan:2016vau}}  \\ 
            \cline{2-4}
			& ATLAS $\mu (h \to Z \gamma)$ & \multirow{1}{*}{1} & \multirow{1}{*}{ Figure~1 of Ref.~\cite{Aad:2015gba}}  \\
			\hline
			\multirow{4}{*}{13 TeV ATLAS} &  $\mu ( h \to Z \gamma )$ at 139 $\ifb$ & 1 &  \cite{Aad:2020plj} \\
			& $\mu ( h \to \mu \mu)$ at 139 $\ifb$ & 1 & \cite{Aad:2020xfq} \\
			Run-II data  & $\mu(h \to \tau \tau)$ at 139 $\ifb$ & 4 & Figure~14 of Ref.~\cite{ATLAS-CONF-2021-044} \\
		    & $\mu( h \to bb)$ in VBF and ${ttH}$ at 139 $\ifb$ & 1+1 & \cite{ATLAS:2020bhl,ATLAS:2020syy}  \\
		    \cline{2-4} 
		    & STXS $h \to \gamma \gamma/ZZ/b \bar{b}$ at $139\ifb$ & 42 & Figures~1 and 2 of Ref.~\cite{ATLAS:2020naq} \\
			& STXS $ h \rightarrow$ $W W$ in ggF, VBF at $139\ifb$ & 11 & Figures~12 and 14 of Ref.~\cite{ATLAS:2021upe} \\
            \cline{2-4}
            & di-Higgs $\mu_{_{HH}}^{b\bar{b}b \bar{b}}$, $\mu_{_{HH}}^{b \bar{b}\tau \bar{\tau}}$, $\mu_{_{HH}}^{b \bar{b} \gamma \gamma}$ & 3  &   \cite{ATLAS:2018dpp,ATLAS:2018rnh,ATLAS:2018uni}  \\
			\hline
			 &  $\mu(h \to b \bar{b})$ in $Vh$ at $35.9/41.5\ifb$ & 2  &   entries from Table~4 of Ref.~\cite{CMS:2020gsy} \\
			 &  $\mu(h \to W W)$ in ggF at $137\ifb$ & 1  & \cite{CMS:2020dvg} \\
			 13 TeV CMS &  $\mu (h\to \mu \mu)$ at  $137\ifb$ & 4  & Figure~11 of Ref.~\cite{CMS:2020xwi} \\
			Run-II data & $\mu (h \to \tau \tau/WW)$ in  $t\bar{t}h$ at $137\ifb$ & 3  & Figure~14 of Ref.~\cite{CMS:2020mpn} \\
			\cline{2-4} 
			 & STXS $h\to WW$ at $137\ifb$ in $Vh$  & 4 &  Table~9 of Ref.~\cite{CMS:2021ixs} 
			\\
			& STXS $h \to \tau \tau$ at  $137\ifb$  & 11 &  Figures~11/12 of Ref.~\cite{CMS:2020dvp} 
			\\
			& STXS $h \to \gamma \gamma$ at  $137\ifb$ & 27 & Table~13 and Figure~21 of Ref.~\cite{CMS:2021kom} 
			\\
			& STXS $h \to ZZ $ at  $137\ifb$ & 18 & Table~6 and Figure~15 of Ref.~\cite{CMS:2021ugl} 
			\\
            \cline{2-4}
            & di-Higgs $\mu_{_{HH}}^{b\bar{b}b \bar{b}}$, $\mu_{_{HH}}^{b \bar{b}\tau \bar{\tau}}$, $\mu_{_{HH}}^{b \bar{b} \gamma \gamma}$ & 3  &   \cite{CMS:2020tkr, CMS:2021ssj,CMS:2017hea}  \\
			\hline
			\multicolumn{2}{|c|}{{\bf{ ATLAS $Zjj$  13 TeV $\Delta \phi_{jj}$}} at $139\ifb$}&12  & Figure~7(d) of Ref.~\cite{ATLAS:2020nzk}  \\
            \hline
		\end{tabular}
	\end{adjustbox}
\end{table}

\begin{table}[ht]
	\centering
	\renewcommand{\arraystretch}{2.0}
	\caption{Top physics observables from Tevatron and LHC Run I included in the fit.}
	\begin{adjustbox}{width=0.9\textwidth}
		\label{tab:obset_top}
		\begin{tabular}{|c|c|c|c|}
			\hline
			\multicolumn{2}{|c|}{Observables} & no. of meas. &  References \\
			\hline
        \multicolumn{2}{|c|}{\bf{Top Data from Tevatron and LHC Run I}} & 82 &  \\ 
			\cline{1-3} 
 Tevatron & 
 forward-backward asymmetry $A_{FB}(m_{t\bar{t}})$ for $\mathrm{t}\overline{\mathrm{t}}$ production  &
$4$ &
 ~\cite{CDF:2017cvy} \\ \hline
\multirow{2}{*}{ATLAS \& CMS}
    & 
    charge asymmetry $A_C(m_{t\bar{t}})$ for $\mathrm{t}\overline{\mathrm{t}}$ production in the $\ell$+jets channel.  &
 $6$ &
 ~\cite{ATLAS:2017gkv} \\ \cline{2-4} 
 & 
   $W$-boson helicity fractions in top decay 
 & $3$ &  ~\cite{CMS:2020ezf} \\ 
  \hline
 \multirow{8}{*}{ATLAS} &  charge asymmetry $A_C(m_{t\bar{t}})$ for $\mathrm{t}\overline{\mathrm{t}}$ production in the dilepton channel  &
 $1$ &
 ~\cite{ATLAS:2016ykb} \\  \cline{2-4} 
     & $\sigma_{t\bar{t}W}, \, \sigma_{t\bar{t}Z}$ &
 $2$ &
 ~\cite{ATLAS:2015qtq} \\  \cline{2-4} 
   & $\tfrac{d\sigma}{dp^T_{t}} ,\quad \tfrac{d\sigma}{d|y_{\bar{t}}|}$  for $t$-channel single-top production
 &
 $4+5$ &
 ~\cite{ATLAS:2017rso} \\ \cline{2-4} 
  &  $\sigma_{tW}$ in the single lepton channel &
 $1$ & ~\cite{ATLAS:2020cwj} \\ \cline{2-4} 
 & $\sigma_{tW}$ in the dilepton channel &
 $1$ & ~\cite{ATLAS:2015igu} \\ \cline{2-4} 
& $s$-channel single-top cross section &
 $1$ & ~\cite{ATLAS:2015jmq} \\ \cline{2-4} 
& $\tfrac{d\sigma}{dm_{t\bar{t}}}$ for $t\bar{t}$ production in the dilepton channel
  &
 $6$ &
 ~\cite{ATLAS:2016pal} \\ \cline{2-4} 
 & $\tfrac{d\sigma}{dp^T_{t}}$  for $t\bar{t}$ production in the $\ell$+jets channel
 & $8$ &
 ~\cite{ATLAS:2015lsn} \\ \hline
  \multirow{10}{*}{CMS}    & $\sigma_{t\bar{t}\gamma}$ in the $\ell+$ jets channel.&
 $1$ &
 ~\cite{CMS:2015uvn} \\ \cline{2-4} 
  & charge asymmetry $A_C(m_{t\bar{t}})$ for $\mathrm{t}\overline{\mathrm{t}}$ production in the dilepton channel.  &
 $3$ &
 ~\cite{CMS:2016ypc} \\ \cline{2-4} 
  & $\sigma_{t\bar{t}W}, \, \sigma_{t\bar{t}Z}$ &
 $2$ &
  ~\cite{CMS:2015uvn} \\ \cline{2-4} 
   & $\sigma_{t\bar{t}\gamma}$ in the $\ell+$ jets channel.&
 $1$ &
  ~\cite{CMS:2017tzb} \\ \cline{2-4} 
   & $s$-channel single-top cross section &
 $1$ &
 ~\cite{CMS:2016xoq} \\ \cline{2-4} 
  &   $\tfrac{d\sigma}{dp^T_{t+\bar{t}}}$ of $t$-channel single-top production
 &
 $6$ & ~\cite{CMS:2014ika} \\ \cline{2-4} 
  & $t$-channel single-top and anti-top cross sections $R_t$. &
 $1$ & ~\cite{CMS:2014mgj} \\ \cline{2-4} 
  & $\sigma_{tW}$ &
 $1$ & ~\cite{CMS:2014fut} \\ \cline{2-4} 
   & $\tfrac{d\sigma}{dm_{t\bar{t}}dy_{t\bar{t}}}$ for $t\bar{t}$ production in the dilepton channel
  & $16$ &
 ~\cite{CMS:2017iqf,CMS:2013hon} \\ \cline{2-4} 
  &  $\tfrac{d\sigma}{dp^T_{t}}$ for $t\bar{t}$ production in the $\ell$+jets channel
  & $8$ &
 ~\cite{CMS:2015rld,CMS:2016csa} \\ \hline
		\end{tabular}
	\end{adjustbox}
\end{table}
 
\begin{table}[ht]
	\centering
	\renewcommand{\arraystretch}{2.0}
	\caption{Top physics observables from LHC Run~II as well as data from Drell-Yan and dijet+photon production included in the analysis. }
	\begin{adjustbox}{width=0.9\textwidth}
		\label{tab:obset_top2_DY_dijet}
		\begin{tabular}{|c|c|c|c|}
			\hline
			\multicolumn{2}{|c|}{Observables} & no. of meas. &  References \\
			\hline
        \multicolumn{2}{|c|}{\bf{Top Data from LHC Run II}} & 55 &  \\ 
			\cline{1-3} 
  \multirow{6}{*}{ATLAS} & $\sigma_{tW}$  &
 $1$ &
 ~\cite{ATLAS:2016ofl} \\ \cline{2-4} 
   & $\sigma_{tZ}$ &
 $1$ &
 ~\cite{ATLAS:2017dsm} \\ \cline{2-4} 
  & $\sigma_{t+\bar{t}}, \, R_t$ for $t$-channel single-top and anti-top cross sections&
 1+1 &~\cite{ATLAS:2016qhd} \\ \cline{2-4} 
   &   charge asymmetry $A_C(m_{t\bar{t}})$ for $\mathrm{t}\overline{\mathrm{t}}$ production &
 $5$ &
 ~\cite{ATLAS:2019czt} \\ \cline{2-4} 
  &  $\sigma_{t\bar{t}W}, \, \sigma_{t\bar{t}Z}$  &
 $2$ &
 ~\cite{ATLAS:2019fwo} \\ \cline{2-4} 
   &  $\tfrac{d\sigma}{dp^T_{\gamma}}$ for  $t\bar{t}\gamma$ production
 &
 $11$ &
 ~\cite{ATLAS:2020yrp} \\ \hline
  \multirow{7}{*}{CMS} & $\sigma_{tW}$ &
 $1$ &
 ~\cite{CMS:2018amb} \\ \cline{2-4} 
   & $\sigma_{tZ}$ in the $Z\to\ell^+\ell^-$ channel  &
 $1$ &
 ~\cite{CMS:2018sgc} \\ \cline{2-4} 
 & 
 $\tfrac{d\sigma}{dp^T_{t+\bar{t}}}$ and $R_t\left(p^T_{t+\bar{t}}\right)$ for $t$-channel single-top quark production &
 $5+5$ &
 ~\cite{CMS:2019jjp} \\ \cline{2-4} 
   & $\tfrac{d\sigma}{dm_{t\bar{t}}}$  for $t\bar{t}$ production in the dilepton channel
 &
 $6$ &
 ~\cite{CMS:2018fks}  \\ \cline{2-4} 
 & $\tfrac{d\sigma}{dm_{t\bar{t}}}$ for $t\bar{t}$ production in the $\ell+$jets channel
&
 $15$ &
 \cite{CMS:2021fhl}  \\ \cline{2-4} 
 & $\sigma_{t\bar{t}W}$ &
 $1$ &
 ~\cite{CMS:2017ugv} \\ \cline{2-4} 
 & $\tfrac{d\sigma}{dp^T_{Z}}$ for $t\bar{t}Z$ production  &
 $4$ &
 ~\cite{CMS:2019too} \\ 
 \hline
     			\multicolumn{2}{|c|}{\bf{Drell-Yan}} & 109 &  \\ 
			\cline{1-3}
			\multirow{3}{*}{13 TeV } & CMS $e^+ e^-$, $m_{ee}$ & 61 (up to 3~TeV) & Figure~2 of~\cite{CMS:2021ctt}  \\
			\cline{2-4}
			& CMS $\mu^+ \mu^-$, $m_{\mu \mu}$&  34 (up to 3~TeV) &  Figure~2 of~\cite{CMS:2021ctt} \\ 
            \cline{2-4}
			& ATLAS $\tau^+\tau^-$, $m_T^{\text{tot}}$ & 14 (up to 3~TeV) & Figure~1 of~\cite{ATLAS:2020zms} \\
   \hline
			\multicolumn{2}{|c|}{\bf{Dijets+photon}} & 26 &  \\ 
			\cline{1-3}
			\multirow{1}{*}{13 TeV } & ATLAS $\frac{d N_{\text{evt}} }{dm_{jj}}$ for $pp\to jj\gamma+X$& 26 (from 500~GeV) & Figure~1 of~\cite{ATLAS:2019itm}  \\
 \hline
		\end{tabular}
	\end{adjustbox}
\end{table}

\begin{table}[h]
	\centering
    \vspace{-1cm}
	\renewcommand{\arraystretch}{2.0}
    \caption{EWPO, PVE, lepton scattering and flavour observables included in the fit.}
	\begin{adjustbox}{width=0.9\textwidth}
		\label{tab:obset_PVE_flavour}
		\begin{tabular}{|c|c|c|c|}
			\hline
			\multicolumn{2}{|c|}{Observables} & no. of measurements	 &  References \\
			\hline
            \multicolumn{2}{|c|}{\bf{Electroweak Precision Observables (EWPO)}}& 13 & Ref.~\cite{ALEPH:2005ab} \\
			\hline
			\multicolumn{2}{|c|}{\bf{PVE and lepton scattering}} & 163 &  \\ 
			\cline{1-3}
			\multirow{4}{*}{PVE} & $Q_W^{\text{Cs}}$ & 1 & \cite{ParticleDataGroup:2016lqr}  \\
			\cline{2-4}
			& $Q_W^{\text{p}}$ & 1 & \cite{Qweak:2013zxf}  \\ 
            \cline{2-4}
			& $A_{1,2}^{\text{PVDIS}}$ & 2 & \cite{PVDIS:2014cmd} \\
            \cline{2-4}
			& SAMPLE & 1 & \cite{Beise:2004py} \\
			\hline
			\multirow{6}{*}{lepton scattering} & $\nu_\mu\nu_\mu e e$ & 2 & \cite{ParticleDataGroup:2016lqr}  \\
            \cline{2-4}
			& $P_\tau,\,A_P$ & 2 & \cite{VENUS:1997cjg}\\
			\cline{2-4}
			& $g_{AV}^{ee}$ in $e^-e^-\to e^-e^-$ & 1 & \cite{ParticleDataGroup:2016lqr}  \\ 
            \cline{2-4}
			&  $A_{\text{FB}}^{\mu,\tau}$ in $e^+e^-\to l^+ l^-$ & 24 & \cite{ ALEPH:2013dgf, Electroweak:2003ram}\\
            \cline{2-4}
			& $\sigma_{\mu,\tau}$ in $e^+e^-\to l^+ l^-$ & 24 & \cite{ALEPH:2013dgf, Electroweak:2003ram}\\
            \cline{2-4}
			& $\frac{d\sigma(ee)}{d\cos{\theta}}$ in $e^+e^-\to l^+ l^-$ & 105 & \cite{ALEPH:2013dgf, Electroweak:2003ram}\\
   			\hline
			\multicolumn{2}{|c|}{\bf{Flavour}} & 37 &  \\ 
			\cline{1-3}
			  \multicolumn{2}{|c|}{Differential  $\text{BR}(B\to K\mu\mu)$ (from $14$ GeV)}  & 3  & \cite{LHCb:2014cxe}  \\
            \multicolumn{2}{|c|}{Differential  $\text{BR}(B\to K^* \mu\mu)$ (from $14$ GeV)}  & 3  & \cite{CMS:2015bcy}  \\
            \multicolumn{2}{|c|}{Differential  $\text{BR}(\Lambda_b\to \Lambda \mu\mu)$ (from $15$ GeV)}  & 1  & \cite{LHCb:2015tgy}  \\
	\cline{1-4}
			\multicolumn{2}{|c|}{$\text{BR}(B\to X_s\mu\mu,\,\mu\mu,\,X_s\gamma,\,K^*\gamma,\,K^{(*)}\Bar{\nu}\nu)$} & 5 & \cite{BaBar:2013qry,Greljo:2022jac,Misiak:2017bgg,HFLAV:2014fzu,BaBar:2013npw}\\ 
   \multicolumn{2}{|c|}{$\text{BR}(B_s\to \mu\mu,\,\phi\gamma$)} & 2 & \cite{Greljo:2022jac,Belle:2014sac}
    \\ 
    \multicolumn{2}{|c|}{$\text{BR}(K^+\to\mu^+\nu_\mu,)$}
    & 1 & \cite{ParticleDataGroup:2022pth} 
    \\ 			
    \cline{1-4}
   \multicolumn{2}{|c|}{ $R_K$ and $R_K^*$}
    & 4 & \cite{LHCb:2022vje} 
   \\ 			
   \cline{1-4}
\multicolumn{2}{|c|}{B meson mixing observables}
    & 2 & \cite{HFLAV:2016hnz} \\ 
	\cline{1-4}
		\multicolumn{2}{|c|}{Angular observables in $B\to K^*\mu\mu$ and $\Lambda_b\to\Lambda\mu\mu$ (from $15$ GeV)}  & 16  & \cite{LHCb:2020lmf,LHCb:2015tgy} \\
     \hline
		\end{tabular}
	\end{adjustbox}
\end{table}

\FloatBarrier
\section{Numerical results}
\label{app:num_res}
\begin{figure}[H]
    \centering
    \includegraphics[scale=0.75]{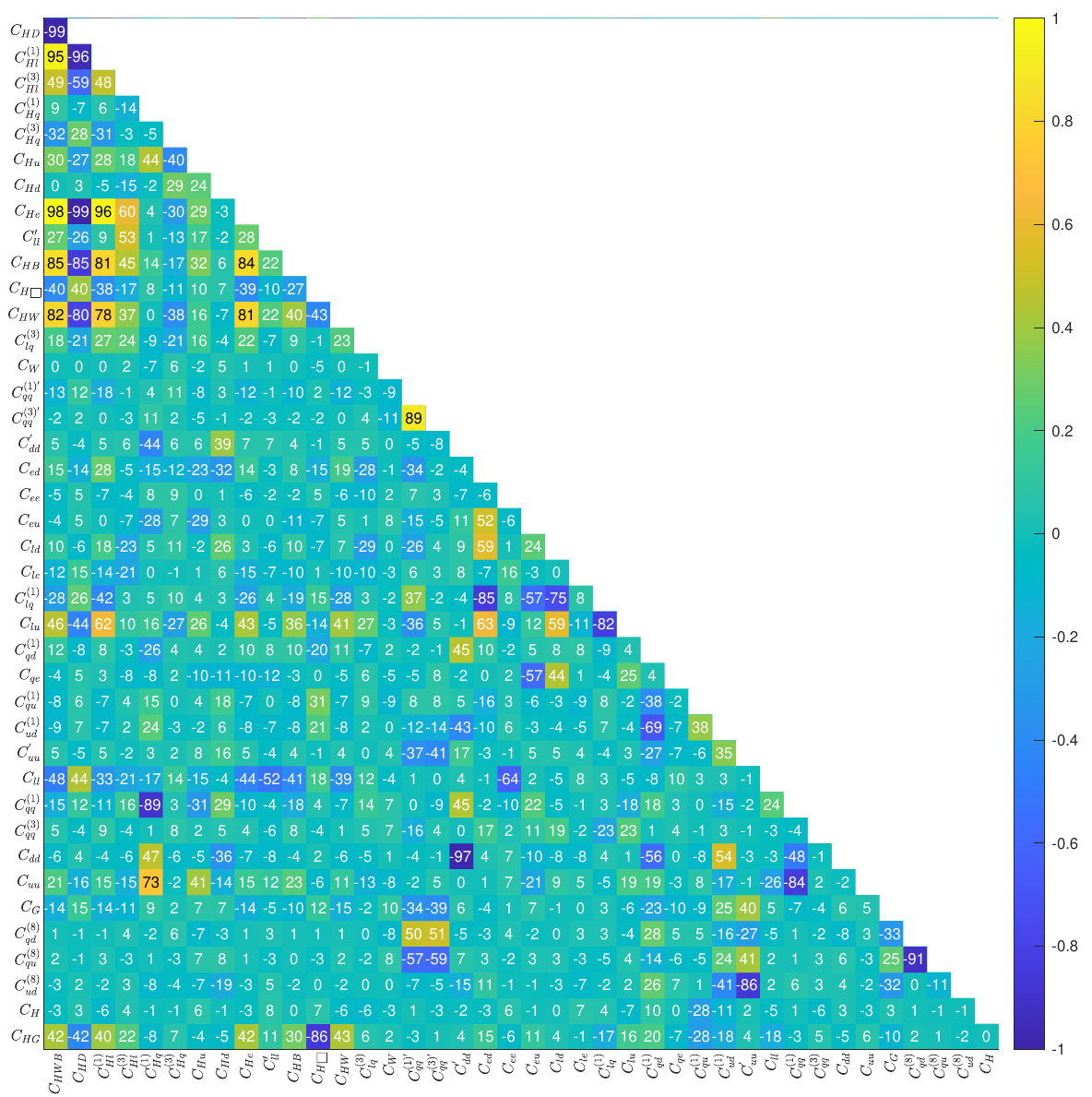}
    \caption{Correlation matrix of the global analysis including partial NLO SMEFT predictions. The numbers in the matrix correspond to the correlations in percent. }
    \label{fig:corr_mat}
\end{figure}

We present the correlation matrix of our global analysis including partial NLO SMEFT predictions in Figure~\ref{fig:corr_mat}. 
Numerical values for the fit results of the single-parameter fit as well as the global fit can be found in Table~\ref{tab:fits_numerical}.

\begin{table}[h]
    \normalsize
    \centering
    \vspace{-0.9cm}
    \begin{tabular}{lcccc}
    \toprule
    & \multicolumn{2}{c}{LO SMEFT predictions} & \multicolumn{2}{c}{partial NLO SMEFT predictions} \\
        coefficient & single 95\%CL limit & global 95\%CL limit  & single 95\%CL limit & global 95\%CL limit  \\
\hline 
$C_{HWB}$ & $[ -0.004  , \,  0.002 ]$ & $[ -0.12  , \,  0.15 ]$ & ${\color{gray}[ -0.004  , \,  0.002 ]}$ & $[ -0.1  , \,  0.18 ]$ \\
$C_{HD}$ & $[ -0.022  , \,  0.003 ]$ & $[ -0.39  , \,  0.26 ]$ & ${\color{gray}[ -0.022  , \,  0.003 ]}$ & $[ -0.43  , \,  0.23 ]$ \\
$C_{Hl}^{(1)}$ & $[ -0.006  , \,  0.011 ]$ & $[ -0.06  , \,  0.11 ]$ & ${\color{gray}[ -0.006  , \,  0.011 ]}$ & $[ -0.04  , \,  0.15 ]$ \\
$C_{Hl}^{(3)}$ & $[ -0.01  , \,  0.003 ]$ & $[ -0.049  , \,  0.063 ]$ & ${\color{gray}[ -0.01  , \,  0.003 ]}$ & $[ -0.062  , \,  0.051 ]$ \\
$C_{Hq}^{(1)}$ & $[ -0.035  , \,  0.014 ]$ & $[ -0.096  , \,  0.055 ]$ & ${\color{gray}[ -0.035  , \,  0.014 ]}$ & $[ -0.14  , \,  0.23 ]$ \\
$C_{Hq}^{(3)}$ & $[ -0.01  , \,  0.013 ]$ & $[ -0.036  , \,  0.048 ]$ & ${\color{gray}[ -0.01  , \,  0.013 ]}$ & $[ -0.04  , \,  0.045 ]$ \\
$C_{Hu}$ & $[ -0.048  , \,  0.04 ]$ & $[ -0.22  , \,  0.12 ]$ & ${\color{gray}[ -0.048  , \,  0.04 ]}$ & $[ -0.17  , \,  0.22 ]$ \\
$C_{Hd}$ & $[ -0.094  , \,  0.016 ]$ & $[ -0.65  , \,  0.16 ]$ & ${\color{gray}[ -0.094  , \,  0.016 ]}$ & $[ -0.56  , \,  0.34 ]$ \\
$C_{He}$ & $[ -0.012  , \,  0.009 ]$ & $[ -0.14  , \,  0.19 ]$ & ${\color{gray}[ -0.012  , \,  0.009 ]}$ & $[ -0.13  , \,  0.21 ]$ \\
$C_{ll}^\prime$ & $[ -0.004  , \,  0.017 ]$ & $[ -0.071  , \,  0.018 ]$ & ${\color{gray}[ -0.004  , \,  0.017 ]}$ & $[ -0.075  , \,  0.016 ]$ \\
\hline
$C_{ll}$ & $[ -0.011  , \,  0.048 ]$ & $[ -0.0  , \,  0.22 ]$ & ${\color{gray}[ -0.011  , \,  0.048 ]}$ & $[ -0.01  , \,  0.21 ]$ \\
$C_{ee}$ & $[ -0.022  , \,  0.04 ]$ & $[ -0.16  , \,  0.02 ]$ & ${\color{gray}[ -0.022  , \,  0.04 ]}$ & $[ -0.16  , \,  0.01 ]$ \\
$C_{le}$ & $[ -0.028  , \,  0.027 ]$ & $[ -0.026  , \,  0.039 ]$ & ${\color{gray}[ -0.028  , \,  0.027 ]}$ & $[ -0.027  , \,  0.039 ]$ \\
\hline
$C_{HB}$ & $[ -0.003  , \,  0.002 ]$ & $[ -0.03  , \,  0.067 ]$ & ${\color{gray}[ -0.003  , \,  0.002 ]}$ & $[ -0.022  , \,  0.077 ]$ \\
$C_{H \square}$ & $[ -1.0  , \,  -0.2 ]$ & $[ -1.4  , \,  0.3 ]$ & ${\color{gray}[ -1.0  , \,  -0.2 ]}$ & $[ -1.3  , \,  0.5 ]$ \\
$C_{HW}$ & $[ -0.009  , \,  0.007 ]$ & $[ -0.19  , \,  0.12 ]$ & ${\color{gray}[ -0.009  , \,  0.007 ]}$ & $[ -0.18  , \,  0.13 ]$ \\
$C_H$ & $[ -9.5  , \,  7.7 ]$ & $[ -11  , \,  6 ]$ & ${\color{gray}[ -9.5  , \,  7.7 ]}$ & $[ -13  , \,  4 ]$ \\
$C_{HG}$ & $[ -0.004  , \,  -0.0 ]$ & $[ -0.004  , \,  0.004 ]$ & ${\color{gray}[ -0.004  , \,  -0.0 ]}$ & $[ -0.005  , \,  0.003 ]$ \\
$C_G$ & $[ -0.36  , \,  0.16 ]$ & $[ -0.44  , \,  0.19 ]$ & ${\color{gray}[ -0.36  , \,  0.16 ]}$ & $[ -0.53  , \,  0.2 ]$ \\
$C_W$ & $[ -0.17  , \,  0.36 ]$ & $[ -0.17  , \,  0.36 ]$ & $[ -0.19  , \,  0.34 ]$ & $[ -0.18  , \,  0.36 ]$ \\
\hline
$C_{lq}^{(1)}$ & $[ -0.005  , \,  0.018 ]$ & $[ -0.67  , \,  0.08 ]$ & $[ -0.004  , \,  0.019 ]$ & $[ -0.61  , \,  0.12 ]$ \\
$C_{lq}^{(3)}$ & $[ -0.006  , \,  0.002 ]$ & $[ -0.09  , \,  0.11 ]$ & ${\color{gray}[ -0.006  , \,  0.002 ]}$ & $[ -0.09  , \,  0.11 ]$ \\
$C_{ed}$ & $[ -0.04  , \,  0.008 ]$ & $[ -0.17  , \,  0.8 ]$ & ${\color{gray}[ -0.04  , \,  0.008 ]}$ & $[ -0.27  , \,  0.65 ]$ \\
$C_{eu}$ & $[ -0.005  , \,  0.011 ]$ & $[ -0.29  , \,  0.34 ]$ & ${\color{gray}[ -0.005  , \,  0.011 ]}$ & $[ -0.35  , \,  0.25 ]$ \\
$C_{ld}$ & $[ -0.039  , \,  0.043 ]$ & $[ -0.3  , \,  1.8 ]$ & ${\color{gray}[ -0.039  , \,  0.043 ]}$ & $[ -0.2  , \,  1.9 ]$ \\
$C_{lu}$ & $[ -0.007  , \,  0.025 ]$ & $[ -0.2  , \,  1.3 ]$ & $[ -0.008  , \,  0.024 ]$ & $[ -0.2  , \,  1.3 ]$ \\
$C_{qe}$ & $[ -0.016  , \,  0.019 ]$ & $[ -0.13  , \,  0.79 ]$ & ${\color{gray}[ -0.016  , \,  0.019 ]}$ & $[ -0.12  , \,  0.78 ]$ \\
\hline
$C_{qd}^{(1)}$ & $[ -34  , \,  23 ]$ & $[ -1760  , \,  1090 ]$ & $[ -0.3  , \,  1.7 ]$ & $[ -3.0  , \,  5.2 ]$ \\
$C_{qu}^{(1)}$ & $[ -4.9  , \,  7.3 ]$ & $[ -380  , \,  340 ]$ & $[ 0.0  , \,  0.6 ]$ & $[ -0.2  , \,  1.8 ]$ \\
$C_{ud}^{(1)}$ & $[ -7.2  , \,  4.7 ]$ & $[ -520  , \,  370 ]$ & $[ -1.8  , \,  0.7 ]$ & $[ -7.4  , \,  3.9 ]$ \\
$C_{qq}^{(1)}$ & $[ -0.13  , \,  0.19 ]$ & $[ -12  , \,  1 ]$ & $[ -0.15  , \,  0.16 ]$ & $[ -6.7  , \,  3.4 ]$ \\
$C_{qq}^{(1)\prime}$ & $[ -0.039  , \,  0.018 ]$ & $[ -0.07  , \,  0.39 ]$ & ${\color{gray}[ -0.039  , \,  0.018 ]}$ & $[ -0.06  , \,  0.42 ]$ \\
$C_{qq}^{(3)}$ & $[ -0.021  , \,  0.04 ]$ & $[ -0.01  , \,  0.11 ]$ & ${\color{gray}[ -0.021  , \,  0.04 ]}$ & $[ -0.01  , \,  0.1 ]$ \\
$C_{qq}^{(3)\prime}$ & $[ -0.017  , \,  0.029 ]$ & $[ -0.05  , \,  0.35 ]$ & ${\color{gray}[ -0.017  , \,  0.029 ]}$ & $[ -0.05  , \,  0.38 ]$ \\
$C_{uu}$ & $[ -0.14  , \,  0.22 ]$ & $[ -2  , \,  15 ]$ & $[ -0.18  , \,  0.17 ]$ & $[ -0.9  , \,  8.8 ]$ \\
$C_{uu}^{\prime}$ & $[ -0.15  , \,  0.07 ]$ & $[ -1.8  , \,  0.3 ]$ & ${\color{gray}[ -0.15  , \,  0.07 ]}$ & $[ -2.3  , \,  0.4 ]$ \\
$C_{dd}$ & $[ -1.8  , \,  2.7 ]$ & $[ -210  , \,  70 ]$ & $[ -1.9  , \,  2.6 ]$ & $[ -190  , \,  30 ]$ \\
$C_{dd}^{\prime}$ & $[ -1.2  , \,  1.8 ]$ & $[ -62  , \,  82 ]$ & ${\color{gray}[ -1.2  , \,  1.8 ]}$ & $[ -19  , \,  84 ]$ \\
$C_{qd}^{(8)}$ & $[ -0.8  , \,  0.21 ]$ & $[ -3.6  , \,  2.7 ]$ & ${\color{gray}[ -0.8  , \,  0.21 ]}$ & $[ -3.8  , \,  3.6 ]$ \\
$C_{qu}^{(8)}$ & $[ -0.18  , \,  0.06 ]$ & $[ -0.7  , \,  0.51 ]$ & ${\color{gray}[ -0.18  , \,  0.06 ]}$ & $[ -0.86  , \,  0.53 ]$ \\
$C_{ud}^{(8)}$ & $[ -0.59  , \,  0.23 ]$ & $[ -4.6  , \,  4.0 ]$ & ${\color{gray}[ -0.59  , \,  0.23 ]}$ & $[ -4.9  , \,  5.8 ]$ \\
\bottomrule
    \end{tabular}
    \caption{Numerical results of the single-parameter and global analyses using purely LO 
    or partial NLO SMEFT predictions. Fit results which do not change between the two setups have been greyed out. }
    \label{tab:fits_numerical}
\end{table}

\FloatBarrier
\section{LEFT Hamiltonians for the relevant flavour observables}
\label{app:LEFT_Hamiltonians}
We present the relevant Hamiltonians for the flavour violating observables in Table~\ref{tab:obset_PVE_flavour} where the corresponding Wilson coefficients are defined at the EW scale.
The Hamiltonian for $d_i \to d_j l^+ l^-$ and $d_i \to d_j \gamma$ transitions is defined as
\begin{equation}
\label{eq:WET1}
\mathcal{H}^{ll}_{\text{eff}}\supset \frac{4G_F}{\sqrt{2}}\left[ - \frac{1}{(4\pi)^2}V_{td_j}^* V_{td_i}\sum_{i=3}^{10}C_i^{d_i d_j} \mathcal{O}^{d_i d_j}_i +\sum_{q=u,c}V_{qd_j}^* V_{qd_i}\, ( C_1^{d_i d_j} \mathcal{O}^{q,\,d_i d_j}_1 + C_2^{d_i d_j} \mathcal{O}^{q,\,d_i d_j}_2 ) \right] \, .
\end{equation}
The operators of relevance to this analysis given by
\begin{align}
\mathcal{O}^{q,\,d_i d_j}_1&= (\bar d_i^\alpha \gamma_\mu P_L q^\beta)(\bar q^\beta \gamma^\mu P_L d_j^\alpha), \nonumber\\
\mathcal{O}^{q,\,d_i d_j}_2 &=(\bar d_i^\alpha \gamma_\mu P_L q^\alpha)(\bar q^\beta \gamma^\mu P_L d_j^\beta), \nonumber\\
\mathcal{O}^{d_i d_j}_7 &=e \, m_{d_i}\left(\bar{d_j}\sigma^{\mu\nu}P_R d_i \right)F_{\mu\nu},\nonumber\\
\mathcal{O}^{d_i d_j}_8 &=g_sm_{d_i}\left(\bar{d_j}\sigma^{\mu\nu}T^AP_R d_i \right)G_{\mu\nu}^A,\nonumber\\
\mathcal{O}^{d_i d_j}_9 &=e^2\left( \bar{d_j}\gamma^{\mu}P_L d_i\right) \left( \bar{\ell}\gamma_{\mu} \ell \right), \nonumber\\
\mathcal{O}^{d_i d_j}_{10} &=e^2 \left( \bar{d_j}\gamma^{\mu}P_L d_i\right) \left( \bar{\ell}\gamma_{\mu}\gamma_5 \ell \right) \, ,
\end{align}
where $\alpha$, $\beta$ are colour indices.
Since our fit is linear in the SMEFT Wilson coefficients, the terms appearing in each observable are given by the interference with the SM. Therefore, we can neglect contributions to flavour observables arising from $\mathcal{O}_{3,4,5,6}$ because the matching of the SM onto these LEFT coefficients gives much smaller values with respect to the remaining operators in Equation~\eqref{eq:WET1}.
The LEFT effective Hamiltonian for $d_i \to d_j \bar \nu \nu$ is given by
\begin{equation}
\mathcal{H}^{\nu\nu}_{\text{eff}} \supset -\frac{4G_F}{\sqrt{2}}\frac{1}{(4\pi)^2} \frac{e^2}{\sin^2 \theta_W}V_{td_j}^* V_{td_i} \,C_L^{d_i d_j} \left(\bar d_{j} \gamma^\mu P_L d_i \right) \left(\bar \nu_k \gamma^\mu (1-\gamma^5) \nu_k \right).
\end{equation}
The LEFT Hamiltonian relevant for charged-current semileptonic decays $d_i \to u_j l \bar{\nu}$ is
\begin{align}
\mathcal{H}_{\text{eff}} = \frac{4G_F V_{u_j d_i}}{\sqrt{2}}C_{\pm}\sum_{l} \left( \bar{l}\gamma_\mu P_L \nu_l \right) \left( \bar u_i \gamma^\mu P_L d_j \right) + \text{h.c.} \, .
\end{align}
This process already exists in the SM and we hence get tree-level contributions to  $C_{\pm}$.
Finally, the LEFT Hamiltonian relevant for meson mixing is
\begin{align}
\label{eqn:mixHeff}
\mathcal{H}_{\text{eff}}^{\text{mix}} &\supset \frac{G_F^2 m_W^2}{16\pi^2} \left(\bar{d_j}^\alpha \gamma^\mu P_Ld_i^\alpha \right)(\bar{d_j}^\beta \gamma^\mu P_Ld_i^\beta)\nonumber \\
&\times \left(\lambda_t^2\, C^{d_i d_j}_{1,mix}(x_t)+\lambda_c^2 \,C^{d_i d_j}_{1,mix}(x_c) +2\, \lambda_c \lambda_t\, C^{d_i d_j}_{1,mix}(x_t, x_c)\right), 
\end{align}
where $\alpha$ and $\beta$ are colour indices, $\lambda_k=V_{kd_j}^* V_{kd_i}$ and $x_k = m_k^2/m_W^2$.

\FloatBarrier
\newpage

\bibliographystyle{JHEP}
\bibliography{bibliography}

\end{document}